\documentclass[pre,twocolumn,superscriptaddress,floatfix,graphicx,showpacs]{revtex4-2}

\usepackage{srctex}
\usepackage{hyperref}
\usepackage{natbib}
\usepackage{amsmath,amssymb}

\usepackage{epsfig}\usepackage{graphicx}

\begin{document}




\title{Local structure, thermodynamics, and melting of boron phosphide at high pressures by deep learning-driven \textit{ab initio} simulations}


\author{N.~M.~Chtchelkatchev*}
\affiliation{Vereshchagin Institute for High Pressure Physics, Russian Academy of Sciences,
108840 Troitsk, Moscow, Russia}

\author{R.~E.~Ryltsev}
\affiliation{Institute of Metallurgy of the Ural Branch of the Russian Academy of Sciences, 620016 Ekaterinburg, Russia}
\affiliation{Ural Federal University, 620002 Ekaterinburg, Russia}

\author{M.~V.~Magnitskaya}
\affiliation{Vereshchagin Institute for High Pressure Physics, Russian Academy of Sciences,
108840 Troitsk, Moscow, Russia}

\author{S.~M.~Gorbunov}
\affiliation{Moscow Institute of Physics and Technology, 141701 Dolgoprudny, Russia}

\author{K.~A.~Cherednichenko}
\affiliation{LSPM-CNRS, Universite Sorbonne Paris Nord, Villetaneuse, France}

\author{V.~L.~Solozhenko}
\affiliation{LSPM-CNRS, Universite Sorbonne Paris Nord, Villetaneuse, France}

\author{V.~V.~Brazhkin}
\affiliation{Vereshchagin Institute for High Pressure Physics, Russian Academy of Sciences,
108840 Troitsk, Moscow, Russia}

\begin{abstract}
Boron phosphide (BP) is a (super)hard semiconductor constituted of light elements, which is promising for high demand applications at extreme conditions. The behaviour of BP at high temperatures and pressures is of special interest but is also poorly understood because both experimental and conventional \textit{ab initio} methods are restricted to studying refractory covalent materials. The use of machine learning interatomic potentials  is a revolutionary trend  that gives a unique opportunity for high-temperature study of materials with \textit{ab initio} accuracy. We develop a deep machine learning potential (DP) for accurate atomistic simulations of solid and liquid phases of BP as well as their transformations near the melting line. Our DP provides quantitative agreement with experimental and \textit{ab initio} molecular dynamics data for structural and dynamic properties. DP-based simulations reveal that at ambient pressure tetrahedrally bonded cubic BP crystal melts into an open structure consisting of two interpenetrating sub-networks of boron and phosphorous with different structures. Structure transformations of BP melts under compressing are reflected by the evolution of low-pressure tetrahedral coordination  to high-pressure octahedral coordination. The main contributions to structural changes at low pressures are made by the evolution of medium-range order in B-subnetwork and at high pressures by the change of short-range order in P-sub-network. Such transformations exhibit an anomalous behavior of structural characteristics in the range of 12--15~GPa. DP-based simulations reveal that $T_m(P)$ curve develops a maximum at $P\approx 13$ GPa, whereas experimental studies provide two separate branches of the melting curve, which demonstrate the opposite behaviour. Analysis of the results obtained raise open issues in developing machine learning potentials for covalent materials and stimulate further experimental and theoretical studies of melting behaviour in BP.
\end{abstract}




\maketitle

\section {Introduction \label{Intro}}

The study of high-temperature properties of covalent materials is a challenging task for both experimental and computational methods. From an experimental point of view, determination, for example, melting temperatures in such materials is a challenging task due to their refractoriness~\cite{Solozhenko2015}. Investigation of the properties of melts is an even more complicated task.

One of the obvious solutions is the use of atomistic computer simulations~\cite{Behler2016,Jinnouchi,Behler2021,Vandermause2020,Choudhary2021,Choudhary2022}, which allow studying almost all observable properties even at extreme thermodynamic conditions.  The global issue in such simulations is the accuracy vs efficiency problem. Indeed, highly accurate \textit{ab initio} methods are too costly to solve many practically important problems especially in a wide range of thermodynamical parameters~\cite{Minakov}. Computationally efficient methods of classical molecular dynamics~\cite{Deshchenya2022} often have low accuracy due to the use of empirical interparticle potentials~\cite{Antropov_2020}. The latter problem is especially relevant for covalent materials where strongly non-isotropic interactions are hardly approximated by simple model functions.

Recently, a revolution has occurred in the development and using of machine learning interatomic potentials (MLIPs)~\cite{Behler2011,Behler2017,Wang2018a,Deringer2019,Vandermause2020,Mishin2021}. Such potentials have a flexible functional form (for example, multilayer neural networks), which makes it possible to effectively approximate the potential energy surface in various classes of materials, even very complicated ones. The undeniable primacy of these new MLIP potentials in the ability to quantify the properties of real materials in the ultra wide range of thermodynamic and configuration parameters unlike interaction potentials of previous generations, like, e.g., EAM or MEAM~\cite{Duff2015}. MLIP-based simulations provide nearly \textit{ab initio} accuracy with orders of magnitude less computational cost~\cite{Mishin2021,Deringer2019,Behler2017,Behler2011}.

A promising application of MLIPs is simulation of systems with strongly nonisotopic chemical interactions~\cite{Balyakin2020,Wang2021,Malosso_2022}. Such systems are widely used in practice and often exhibit complex behavior, which is interesting from a fundamental point of view. One of the promising materials of this class is boron phosphide (BP), an equiatomic compound of boron and phosphorus. Under ambient conditions, BP has the structure of cubic sphalerite, also known as zinc-blende (zb) structure (space group F$\overline{4}$3m \#216), where B and P atoms are tetrahedrally coordinated to each other. This zb-BP phase has been found to be stable up to pressures of the megabar range. Another tetrahedral structure of BP, a hexagonal wurtzite-type polymorph w-BP, has also been reported~\cite{Villars1985}. However, there is no literature data on the stability region of this phase.

In the solid state, BP is a hard refractory wide-gap $A^{\mathrm{III}}B^{\mathrm{V}}$ semiconductor~\cite{Stone1960} exhibiting a unique combination of mechanical~\cite{Bushlya2019,Oganov2014}, thermal and electrical~\cite{Solozhenko2015,Kumashiro1989,Stone1960} properties as well as excellent transport characteristics such as thermal conductivity~\cite{Kumashiro1989} and thermoelectric power~\cite{Yugo1980}. These properties make BP a promising material for a wide range of engineering applications~\cite{Woo2016}. In addition to practical interest, BP demonstrates non-trivial behavior, for example, the decrease in the melting point with increasing pressure~\cite{Solozhenko2015}.

Urgent tasks are fabricating new BP-based materials~\cite{Gui2020}, as well as searching for new structural modifications of this system at high pressures. To provide optimization and theoretical background for archiving these tasks, it is necessary to develop reliable interparticle potentials for simulating structural, dynamic, and thermodynamic properties of BP in wide ranges of temperatures and pressures. In this paper, we report developing such a potential based on deep neural networks using the DeePMD package~\cite{PLIMPTON1995,Wang2018}. This deep learning potential (DP) is verified using both \textit{ab initio} molecular dynamics data for structural and dynamic properties. The impact of high pressures on the structure and thermodynamic properties of BP is studied. We also present original experimental data on the melting of BP in the 20--40~GPa pressure range obtained using the diamond anvil cell (DAC) technique.

\section{Methods}
\subsection{Experimental}
Cubic boron phosphide has been studied in the 20--40~GPa pressure range up to 3100~K by angle-dispersive synchrotron X-ray diffraction at BL10XU beamline of SPring-8~\cite{Hirao2020}. The sample (single-crystal or bulk polycrystalline BP) was loaded in a membrane-type diamond anvil cell with large optical aperture and 400-$\mu$m diameter culets. Argon and KCl were used as pressure medium and thermal insulator, respectively, with advantage of chemical inertness with regard to BP melt. High-flux focused monochromatic beam ($\lambda=0.4139$~\AA) from in-vacuum undulator allowed collecting high-quality diffraction patterns in low-exposure time. The diffraction patterns were recorded by XRD0822 (PerkinElmer) flat panel detector and integrated using Fit2D software~\cite{Hammersley1996}. The sample pressure was determined on-line from the shift of the ruby R1 fluorescence line and from thermoelastic equation of state of KCl~\cite{Walker2002}; the pressure uncertainty at high temperature was less than 2~GPa. Homogeneous double-sided sample heating was achieved by use of 100~W SPI fiber laser heating system focused down to 20~$\mu$m. Double-sided temperature measurements were accomplished through standard grey body radiation measurement via a HRS300 spectrograph system (Princeton Instruments). Temperature was measured before and after each acquisition of diffraction pattern, and the average values were used. Above 2000~K the temperature uncertainty is estimated as $\pm150$~K.

\subsection{Technique of density functional calculations}

\begin{figure*}
  \centering
  \includegraphics[width=0.78\columnwidth]{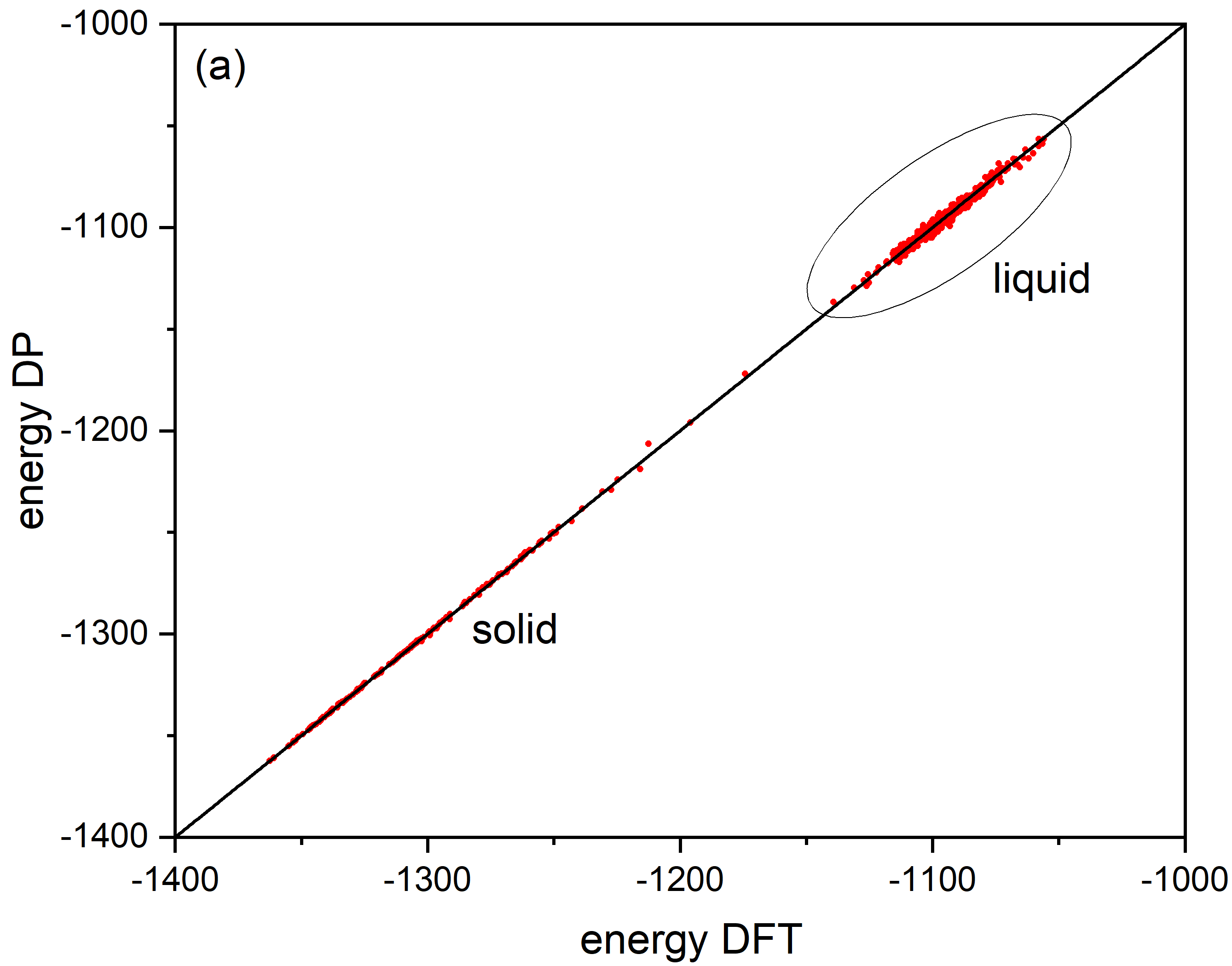} \includegraphics[width=0.74\columnwidth]{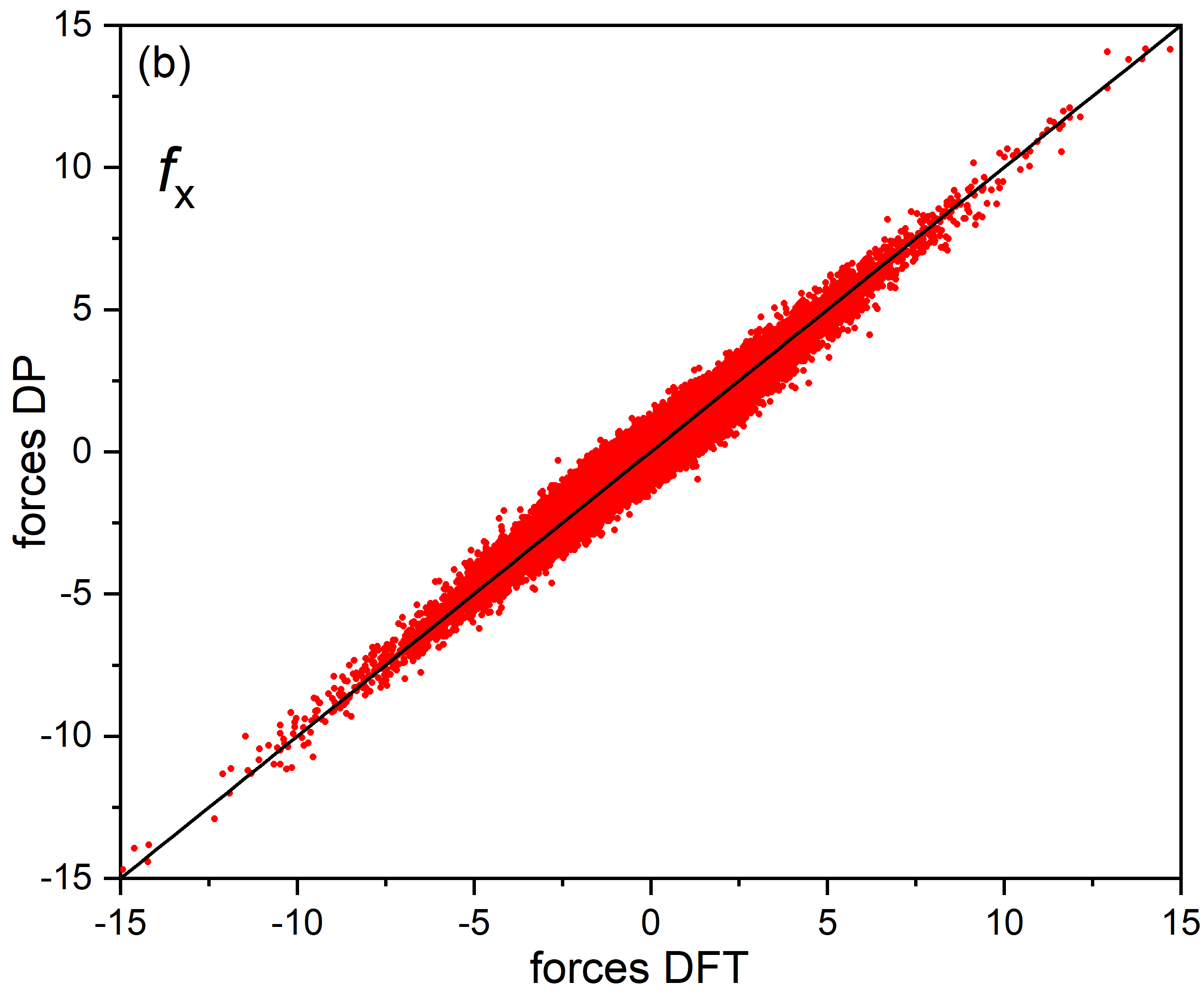}
\hspace{4pt}
  \caption{DP vs DFT curves for energies (a) and forces (b) in boron phosphide.}
  \label{fig:ener-for}
\end{figure*}

We performed \textit{ab initio} simulations within density functional theory (DFT) as implemented in the Vienna \textit{ab initio} simulation package (VASP)~\cite{Kresse1996}. The projector-augmented-wave (PAW) pseudopotential method~\cite{Blochl1994,Kresse1999} with the semilocal PBE–GGA version~\cite{Perdew1996} of the exchange-correlation functional were utilized. The plane-wave kinetic cut-off energy was set to 600 eV and fine Monkhorst-Pack~\cite{Monkhorst1976} \textbf{k}-point grids for sampling the Brillouin zone (BZ) with a reciprocal-space resolution of 0.08~\AA$^{–1}$ were used. The calculations were started from the experimental lattice parameters, then by means of cell relaxation, the equilibrium lattice parameters corresponding to zero pressure were found. The relaxation of atomic coordinates was continued, until the residual atomic forces were converged down to 5 meV/A. The total energy convergence was better than $10^{–6}$~eV/cell.

Melting of BP was simulated using the \textit{ab initio} molecular dynamics (AIMD) as implemented in the VASP package. We considered elementary cells of 512 atoms with periodic boundary conditions, in the $\Gamma$ point. The most disordered initial configurations (special quasirandom structures) were created using the USPEX code~\cite{Oganov2006,Lyakhov2013}. The equilibrium configuration was achieved in the NPT ensemble for at least 10~ps with a step of 1~fs. The AIMD simulation approach is described in detail in~\cite{Kamaeva2020}.

\section{Development of machine learning interatomic potential}

\begin{figure*}
	\centering
		\includegraphics[scale=0.55]{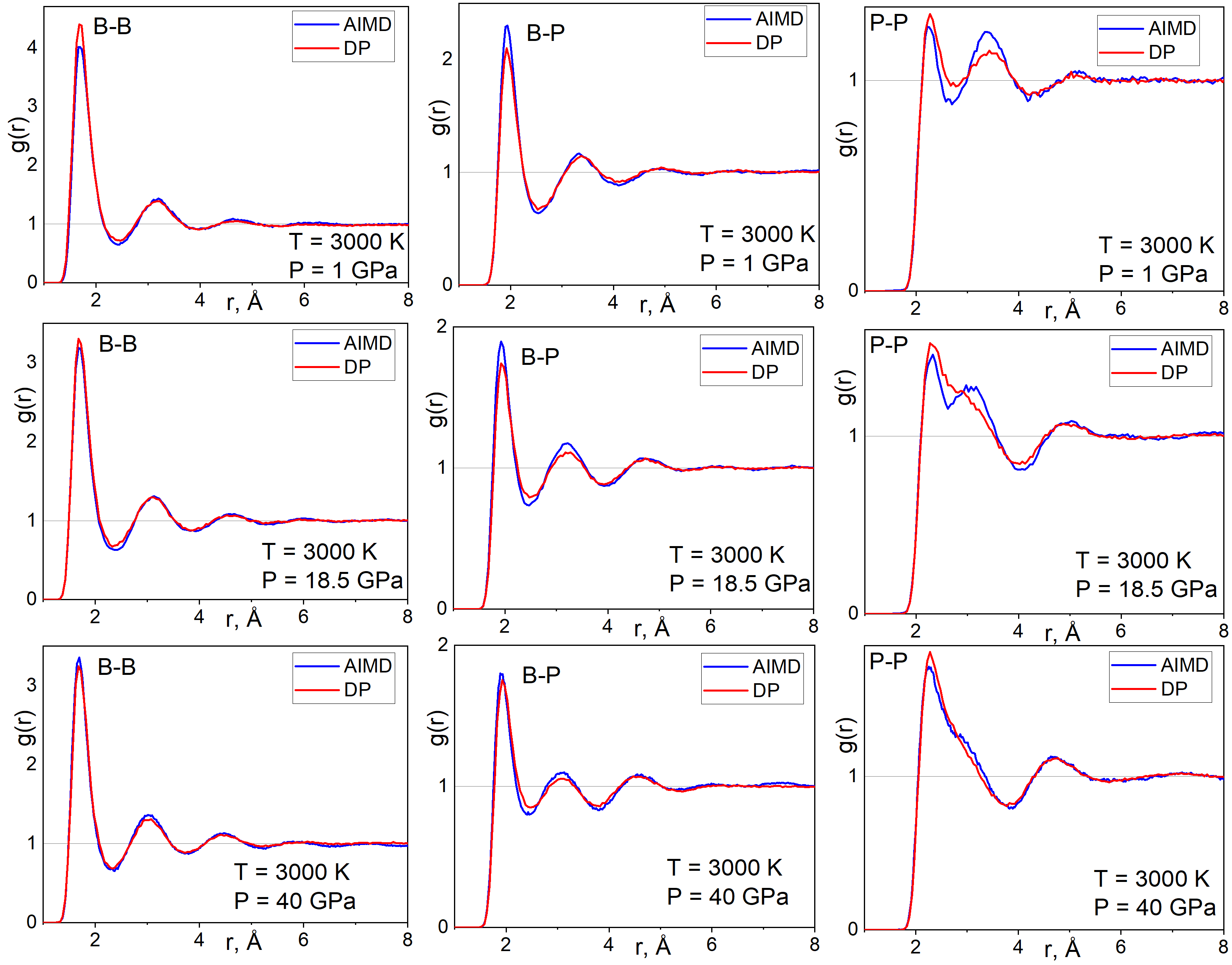}
	\caption{Partial radial distribution functions obtained for BP melt using the DP and AIMD approaches at $T = 3000$~K and $p = 1, 18.5, 40$~GPa. The results of AIMD- and DP-based simulations are shown with blue and red color.}
	\label{fig:rdf}
\end{figure*}

\begin{figure*}
	\centering
		\includegraphics[scale=0.45]{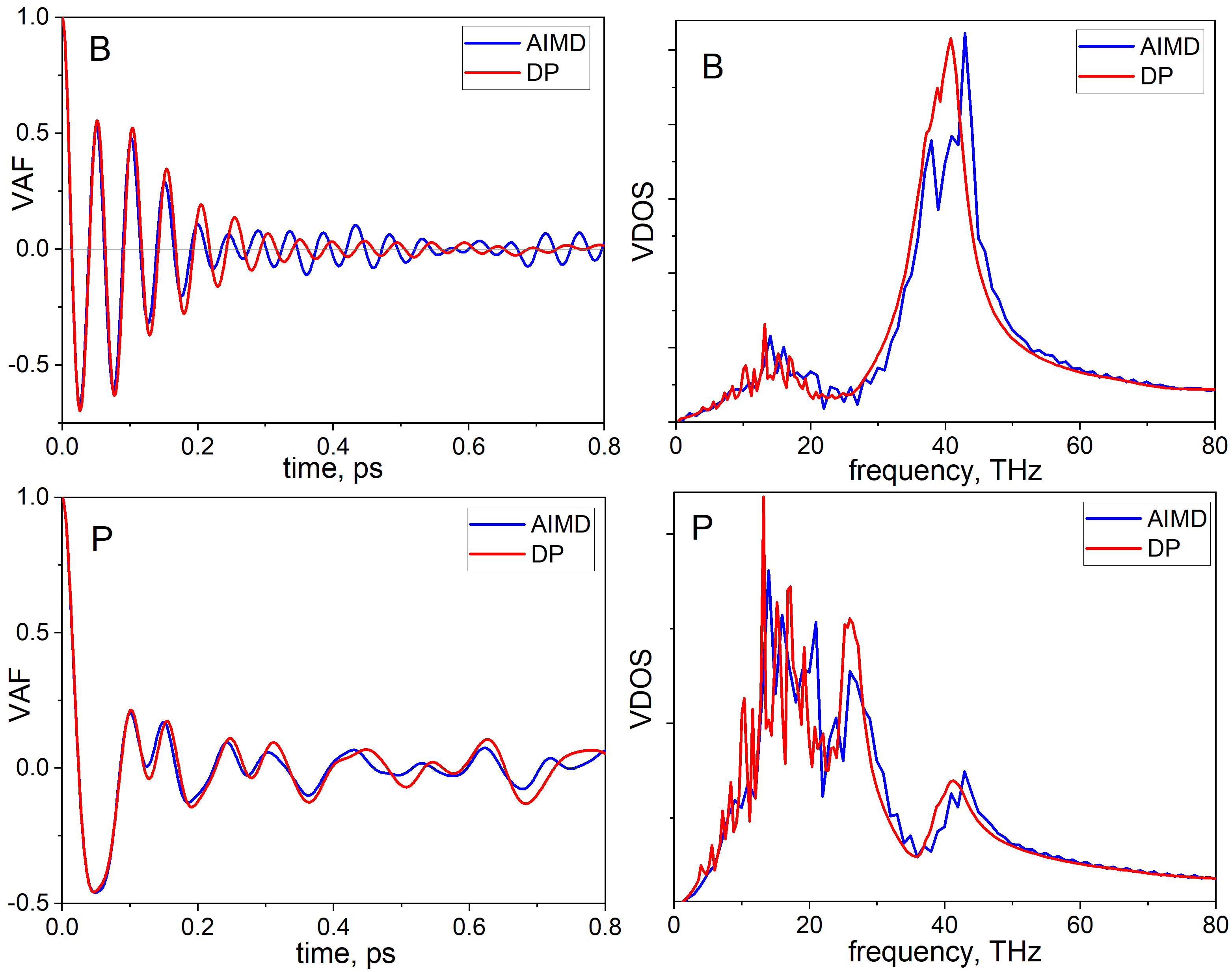}
 	\caption{Velocity autocorrelation functions (left) and corresponding vibrational density of states (right) for solid sphalerite-type BP at $T = 2500$~K and $p = 1$~GPa. Blue lines are the AIMD results and red lines represent DP-based simulations.}
	\label{fig:vaf}
\end{figure*}

\subsection{Generating training dataset}
Our approach for developing deep neural network potentials (DPs) is described in detail in~\cite{ryltsev2022,Kondratyuk2023JML}. The training dataset for DP parametrization was generated using the active learning (AL) procedure in the DPGEN package~\cite{Zhang2020}. During the AL procedure, we trained an ensemble of four DPs with different initializations and performed parallel MD simulations of liquid and crystalline BP phases at various pressures in the range of 0--50~GPa and temperatures in the range of 2000--4000~K. At each MD step, the root-mean-square deviation $\delta$ of forces averaged over ensemble of models was calculated. Configurations for which this deviation exceeded the threshold value $\delta=0.12$--0.2 were selected as candidates for including in the extended dataset. Then, some fixed number of selected configurations were labeled (that means calculation of energies, forces and virials by the DFT method) and included as an extension to the current training dataset. Then, the potential was retrained with an extended dataset, and the simulation process continued with a new version of the potential. This procedure continued in an iterative manner until new configurations for expanding the dataset ceased to be selected for $t_{\mathrm{max}}=25$~ps. As a result, a training dataset of atomic configurations and corresponding values of \textit{ab initio} energies, forces and virials was obtained.

\subsection{Training and verification of the potential}
Using the training dataset described above, we trained several versions of DPs with different values of key hyperparameters, such as the numbers of hidden layers and nodes in both embedding and fitting networks, the cutoff radii etc ~\cite{ryltsev2022}. For all the obtained DPs, the accuracy and the computational efficiency (performance) were evaluated.  We found that all developed models revealed similar accuracy and performance. A model which exhibited slightly better results was chosen for further research. For this best model, the configurations of embedding and fitting nets are respectively $(25,50,100)$ and $(240,240,240)$ and the value of the potential cutoff is 7\AA. The accuracy of this model was determined by the vector ($\delta e$, $\delta f$, $\delta v$) = (6.84, 316, 23) of the average deviations of the values of energy, forces, and virials predicted by the model from the values obtained by DFT method. Here, $\delta e$ and $\delta v$ are presented in units of meV/atom and $\delta f$ in ${\rm meV/\AA}$. The performance of the model is 0.75 ns/day, which was determined as the speed of calculating a 100-step-long trajectory for a system of $N$ = 1728 particles simulated on a single GPU k40m. Note that even the best achieved accuracy is several times less than that usually reported for metals and  binary/ternary metallic alloys, for which the typical force RMSE  $ \delta f < 100$ ${\rm meV/\AA}$ \cite{ryltsev2022}. It underlines the complexity of DP development for covalent materials. Despite a relatively large RMSE, the developed model reveals a pronounced linear correlation between DP and DFT values of energies, forces and virials, see Fig~\ref{fig:ener-for}.


To verify further the developed DP model, it is necessary to compare some of the observed characteristics of the system obtained by DP-based simulations and calculated by the AIMD. For this purpose, partial radial distribution functions (RDF) and velocity autocorrelation functions (VAF) in were chosen. Figure~\ref{fig:rdf} compares the RDF obtained for a boron phosphide melt using DP and AIMD at temperature $T=3000$~K and pressures $p=1, 18.5, 40$~GPa. Figure~\ref{fig:vaf} presents the same comparison for velocity autocorrelation functions and vibrational density of states for BP cubic crystal. It can be seen from the figures that the DP and AIMD calculations demonstrate good quantitative agreement for both structural and dynamical properties.

\begin{figure}
  \centering
  \includegraphics[width=0.99\columnwidth]{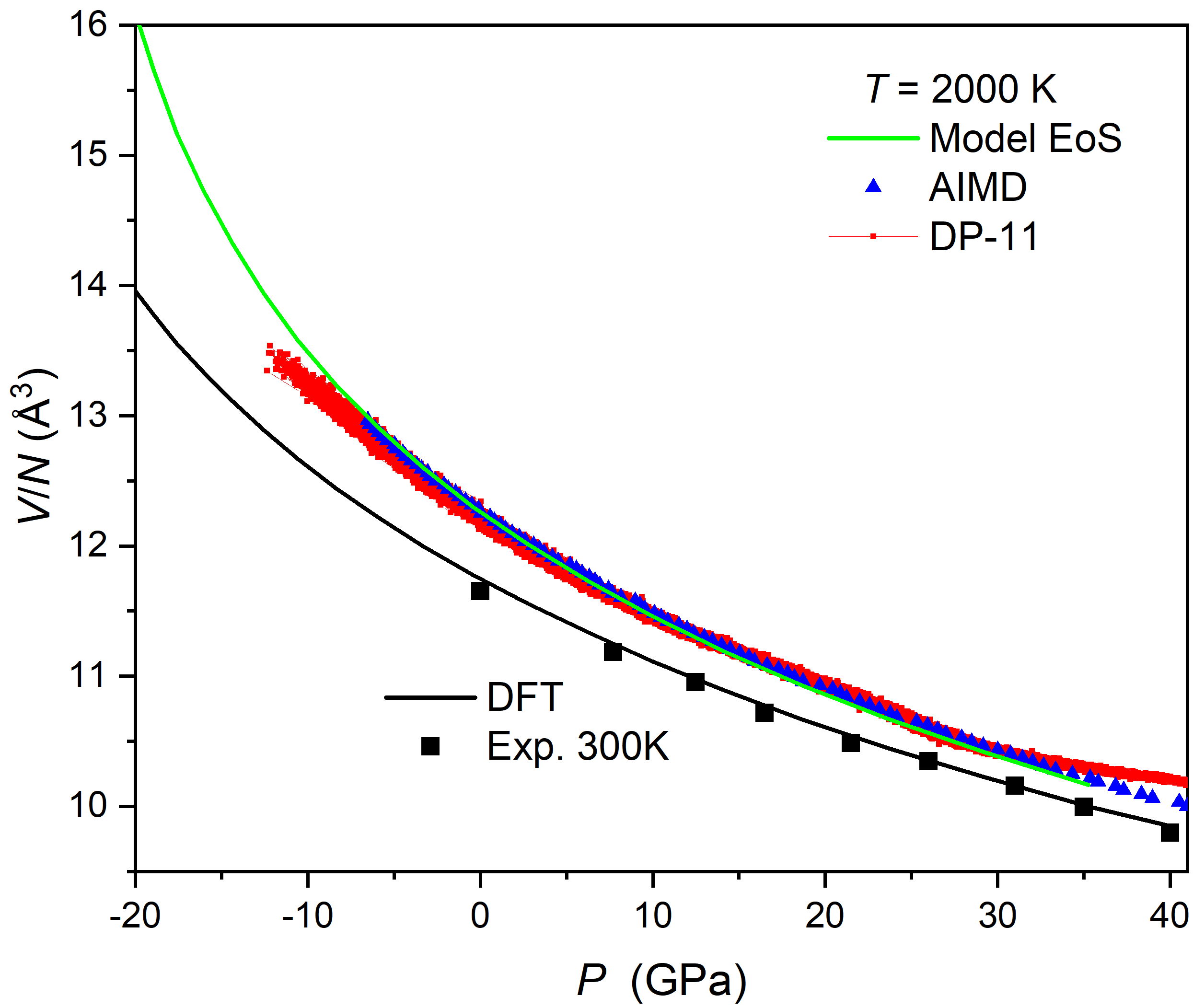}
  \caption{ (a) The $P$--$V$ isotherms for solid and liquid BP as functions of pressure at different temperatures. The volume is normalized on the is the number of atoms $N$.  Black dots represent experimental data for crystal state at $T=300$ K; blue solid lines are the results of AIMD simulations; DP-based simulations are shown with red stars; dashed green line represents the model EoS reported in Ref.~\cite{Kurakevych2015}. (b) Calculated equation of states $\rho(P)$ for  solid (dashed lines) and liquid (solid lines) BP at different temperature demonstrating an intersection points of the curves (density anomaly).}
   \label{fig:isoth-eos}
\end{figure}

Another test of DP predictability was performed on the equation of state $p(V)$ for both solid and liquid phases of BP. In Fig.~\ref{fig:isoth-eos}(a) we present the comparison of  $p$--$V$ isotherms  calculated by AIMD and DP simulations in comparison to experiment and model calculations. We see that calculated isotherms at $T=300$ K are very close to the room-temperature experimental data for single-crystal zb-BP (black dots)~\cite{Oganov2014}. At zero pressure, calculated unit-cell volume, $V_0$, bulk modulus, $B_0$, and its pressure derivative, $B'_0$, were found to be 93.97~\AA$^3$, 163~GPa and 3.7, respectively. This is consistent with the theoretical ($V_0=93.95$~\AA$^3$, $B_0=165$~GPa, $B'_0=3.3$) and experimental ($V_0=93.45$~\AA$^3$, $B_0=174$~GPa, $B'_0=3.2$) values reported in~\cite{Oganov2014}. The results of both AIMD and DP simulations for crystal BP at 2000~K are close to each other and practically coincide with semi-empirical calculations using the model $p$--$V$--$T$ equation of state (EoS) reported in Ref.~\cite{Kurakevych2015} (green dashed line). This model EoS has been developed by combining experimental data and semi-empirical estimations for a large number of different high-pressure-synthesized compounds, including cubic BP. At $T = 2000$ K, the results of DP-based simulations (red stars) deviate slightly from the DFT-based results (blue line) at lower and higher pressures. In the range between $-7$ and 33~GPa, the agreement is good. Good agreement between DP-based and AIMD simulations also revealed for BP melt at $T=3000$ K. It is interesting that calculated $p(V)$ isotherms demonstrate their intersection for liquid and overheated solid curves at certain pressures. This fact is illustrated in Fig.~\ref{fig:isoth-eos}(b) where we show pressure dependencies of the density of liquids and solids at different temperatures. As seen from the picture, the intersection points of liquid and solid curves form a locus of density anomaly on the P-T plot.  We will discuss this effect in the Sec.~\ref{sec:liq-struct} in the context of a maximum on the melting line.

Thus, by analysing the above results for structural, dynamical and thermodynamic properties, we conclude that the developed DP is reliable for simulating properties of both liquid and solid phases of BP at a wide range of temperatures and pressures.

\section{Results}

\subsection{Local structure of BP melt and its evolution under pressure \label{sec:liq-struct}}


\begin{figure*}
	\centering
	\includegraphics[width=0.99\textwidth]{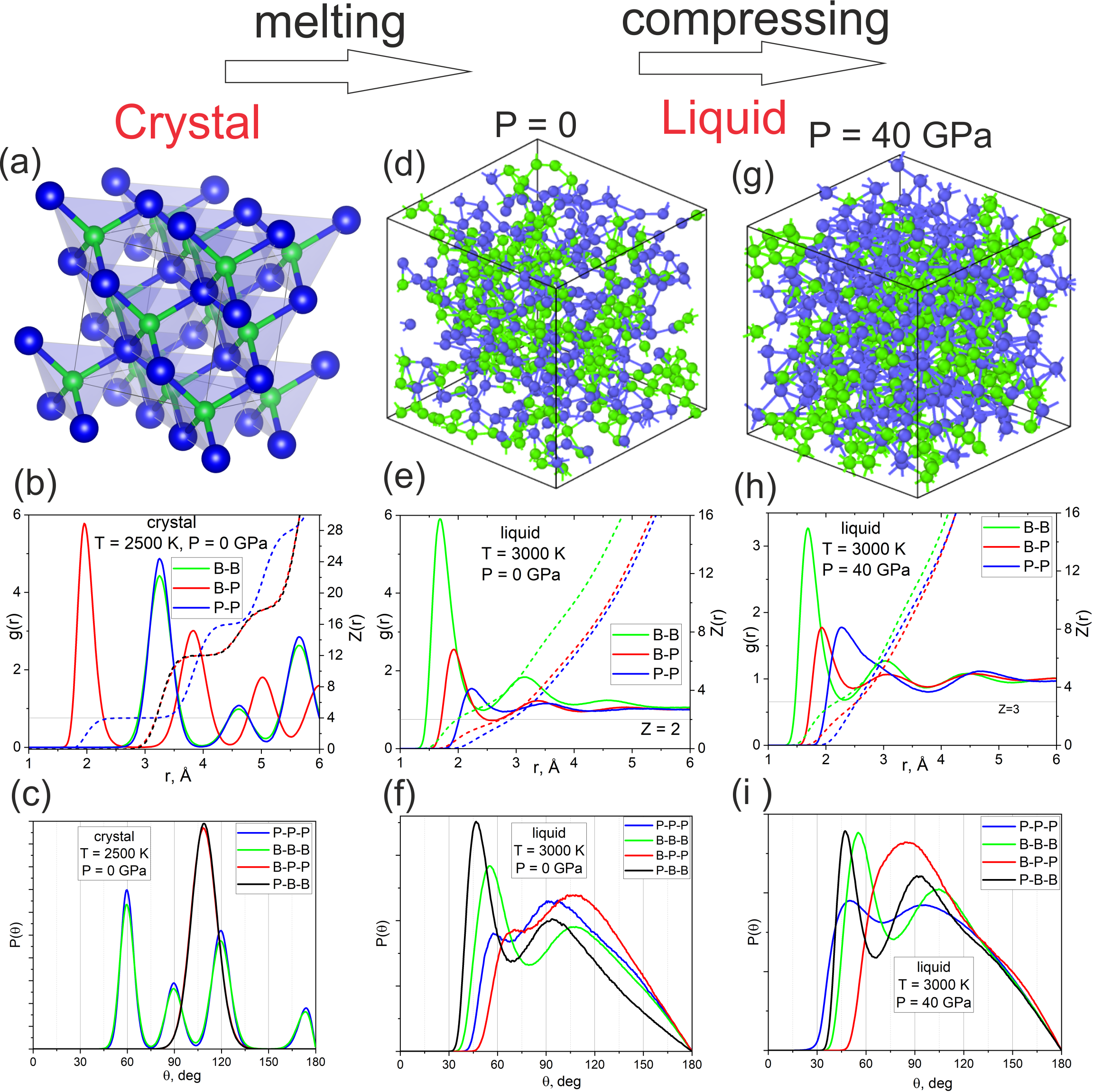}
  \caption{Structure evolution of boron phosphide under melting and compression. Panels (a-c), (e-f) and (g-i) represent respectively snapshots of typical atomic configurations, radial distribution functions $g(r)$ (and their cumulants $Z(r)$) and bond angle distribution functions $P(\theta)$ for crystal BP, liquid BP at $P = 0$ and $P = 40$ GPa.} \label{fig:liq_struct}
\end{figure*}



\begin{figure*}
  \centering
  \includegraphics[width=0.99\textwidth]{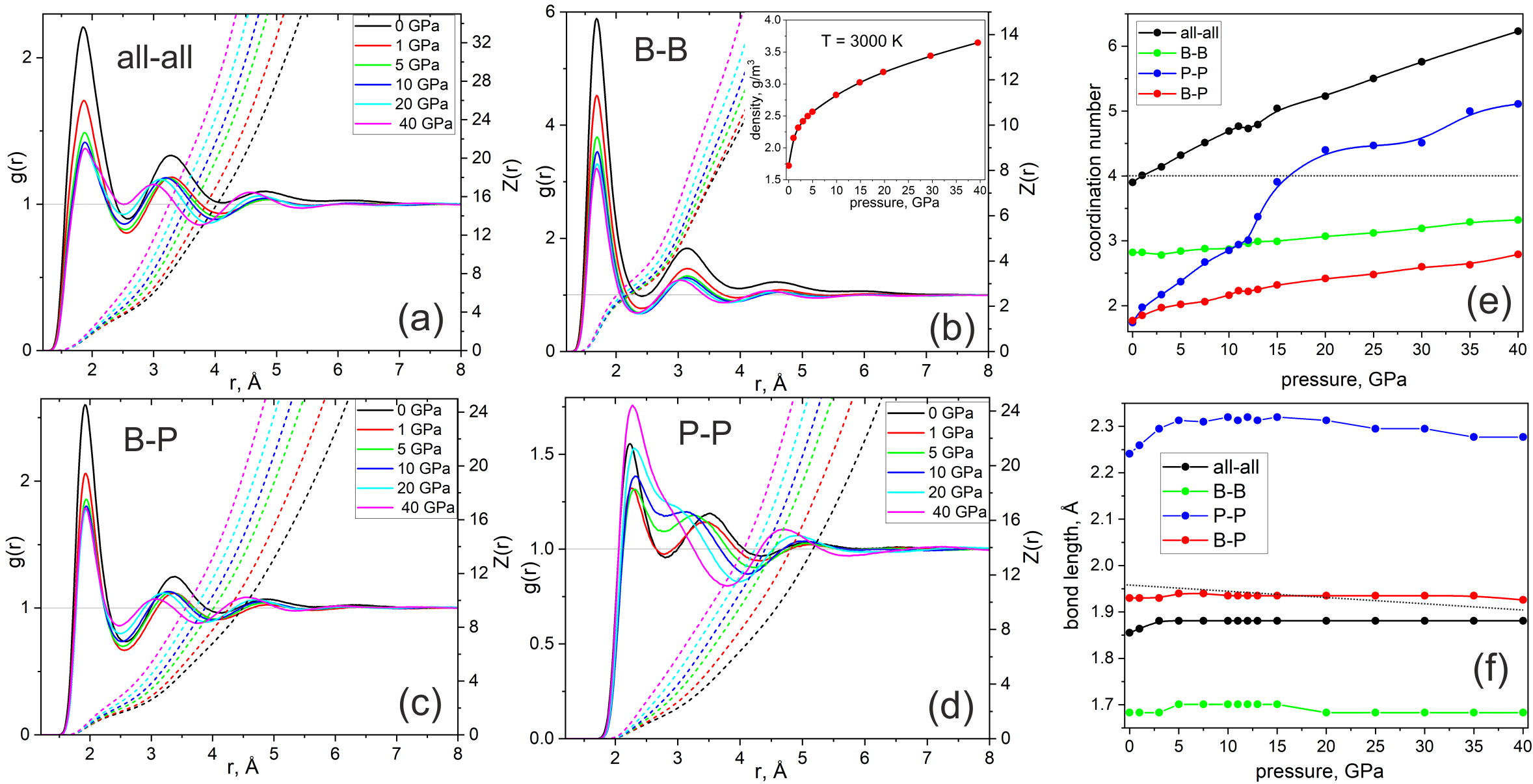}
  \caption{Pressure evolution of RDFs (a-d), total and partial bond lengths (e) and coordination numbers (f) for the BP melt at $T=3000$~K. The dashed black lines in panels (e,f) refers to the crystalline zb-BP. Inset in panel (b) shows pressure dependence of the density at $T=3000$ K.}
    \label{fig:compress}
\end{figure*}





In crystal state at ambient pressure, BP has the structure of cubic sphalerite (space group F$\overline{4}$3m \#216), where B and P atoms are tetrahedrally coordinated to each other, see Fig.\ref{fig:liq_struct}(a).
The calculated partial RDFs show that both B-B and P-P bond lengths are equal to 3.25~\AA~and B-P bond length is 1.95~\AA~ (Fig.\ref{fig:liq_struct}(b)) and coordination numbers for B-P pairs are equal to 4 (see cumulative RDFs $Z(r) = N(<r)$ in Fig.\ref{fig:liq_struct}(b)). Bond angle distribution functions (BADFs) $P(\theta)$ also reflects tetrahedral local order (Fig.\ref{fig:liq_struct}(c)). Indeed, the most probable nearest neighbors angles are equal to 60$^\circ$ for B-B-B and P-P-P triangles and are about 109$^\circ$ for B-P-P and P-B-B triangles, which are in accordance with the geometry of perfect tetrahedron.

The described tetrahedral local structure of BP cubic crystal changes drastically under melting (see Fig.\ref{fig:liq_struct}(d)-(f)). Indeed, while the liquid retains total four-fold coordination, the structure corresponding to individual species is absolutely different from that for crystal state. In particular, B-B and P-P bond lengths, which are equal to 3.25~\AA~ in crystal state, diverge dramatically under melting and become respectively smaller ($r_{\mathrm{B-B}} \approx 1.69$~\AA) and ($r_{\mathrm{P-P}} \approx 2.24$~\AA). Actually, the B-B and P-P distances become very close to the corresponding covalent atomic diameters of boron and phosphorus. In spite of significant shortening of B-B and P-P bond lengths, the system demonstrates drastic decrease in density under melting at ambient pressure, for example, at $T = 3000$ K, $\rho_{\rm cryst} = 2.76$ g/cm$^3$ and $\rho_{\rm liq} = 1.66$ g/cm$^3$. This behaviour can be explained by the fact that liquid BP at ambient pressure forms two interpenetrating networks of boron and phosphide (Fig.\ref{fig:liq_struct}(d)) with sparse areas (voids) providing low density of the whole system. The structures of B and P sub-networks at $P=0$ are different: boron-network tends to form compact clusters with many shared triangles ($Z_{B-B} \approx 3$ and the most pronounced B-B-B BADF peak is $\theta \approx 60^{\circ}$) whereas phosphorous-network is more evenly distributed in space and has strong tendency to form chain-like structures ($Z_{B-B} \approx 2$ and the most pronounced P-P-P BADF peak is $\theta_{B-B} \approx 94^{\circ}$) (see Figs.~(Fig.\ref{fig:liq_struct}(d)-(g)).



To trace changes in the local structure of the BP melts upon compression, we performed simulations up to $P=40$~GPa. In Fig. \ref{fig:liq_struct}(g)-(i)) we demonstrate the characteristics of final state at $P=40$~GPa and Fig.~\ref{fig:compress} shows pressure evolution of RDFs, coordination numbers, bond lengths and density at $T = 3000$~K. We see that total RDF as well as partial RDFs for B-P and B-P pairs demonstrate monotonous decrease in the height of the first peaks and slight shift of their positions (bond lengths) as the pressure increases. The strongest pressure-induced variations are observed for P-P pairs. We see a shift in the positions of the first maximum and the first minimum of $g_{\rm P-P}(r)$, a change in the ratio of heights of the first and second peaks and a strong change of the second peak. As a result, at high pressures, the first and second peaks of $g_{\rm P-P}(r)$ overlap. All this indicates significant changes in the local environment of the P atoms under pressure. Pressure dependencies of coordination numbers $Z(P)$ show a clear trend of their rise, that correlates with a substantial increase in the density (see Fig.~\ref{fig:isoth-eos}(b) and inset in Fig.~\ref{fig:compress}(b)). It is noteworthy that the dependencies $Z_{\mathrm{all-alll}}(P)$ and  $Z_{\mathrm{P-P}}(P)$ demonstrate kinks at $P\approx 12$ GPa, suggesting some drastic changes in the local structure of the melts at this pressure range. However, coordination numbers for P are ill-defined at high pressures due to peak overlapping and so the uncertainty in finding the $Z_{\mathrm{P-P}}$ value is rather large. In any case, the rapid rise of $Z_{\mathrm{P-P}}(p)$ above $\sim 12$~GPa is undoubted. As a result, in the range from 0 to 40~GPa, the low-pressure four-coordinated open structure transforms to a more close-packed structure with coordination close to 6.

Note that for all presented RDFs (especially B-B) we also see strong variations in second and third peaks taking place at the pressure range (0-2) GPa. This indicates changes in the medium-range order, which are mainly related to the collapse of the voids and the formation of more regular and uniform network structure.

Thus, our analyses show that compressing causes significant changes in both short- and medium-range orders of BP melts, especially in local ordering of phosphorus atoms. The data presented in Fig.~\ref{fig:compress}  suggest a structural anomaly in liquid BP located in the pressure range of 12--15~GPa. This conclusion is supported by the behaviour of BP specific volume (density) presented in Fig.~\ref{fig:isoth-eos}. Indeed, the $V(P)$ and $\rho(P)$ curves intersect at pressures which are close to those corresponding to structural anomaly. Below we show that these pressure anomalies of liquid properties correlate with the maximum on the pressure dependence of the melting temperature (see Fig.~\ref{fig:Tm}).


\section{Phase transitions in boron phosphide}

\subsection{Melting curve calculation by thermodynamic integration}
\begin{table*}
\centering
\caption{\label{table-TI}
Numerical integration $\lambda$ values, given in "begin,end,step" notation, used in HTI. First column enumerates potentials between which integration is performed.}
\begin{tabular}{ ccc }
   \hline
	Integration way & cubic BP & liquid BP \\
	\hline
	 Oscillator & [0.0,0.1,0.0125; 0.1,0.2,0.025; & [0.0,0.03,0.003; 0.03,0.06,0.005; \\
     LJ+Oscillator & 0.2,1.0,0.2] & 0.06,0.16,0.01; 0.16,0.3,0.02; 0.3,1.0,0.05] \\
    \hline
	Oscillator+LJ & [0.0,0.05,0.01; 0.05,0.15,0.02; & [0.0,0.006,0.002; 0.006,0.03,0.004;\\
    Oscillator+LJ+DP & [0.0,0.35,0.04; 0.35,1.0,0.065] & 0.03,0.1,0.01; 0.1,0.4,0.03; 0.4,1.0,0.06] \\
   \hline
    Oscillator+LJ+DP &  [0.0,0.75,0.125; 0.75,0.9,0.05; & [0.0,0.75,0.125; 0.75,0.9,0.05; \\
   DP &  0.9,0.96,0.02; 0.96,1.0,0.01] & 0.9,1.0,0.02] \\
   \hline
\end{tabular}
\end{table*}

The melting temperature at a given pressure was found from the equality of the Gibbs free energies ($G$) of the crystalline and liquid phases, with $G$ determined using the thermodynamic integration (TI) method~\cite{Vega2008}.

First, an initial zb-BP supercell of 512~atoms was prepared. Then a larger cell consisting of 1000~atoms was melted by heating to a temperature of 6000~K. The overheated melt obtained in this way was cooled to 3000~K and relaxed in the NVE ensemble. The resulting supercell was used to calculate the liquid phase. The simulations were performed using the LAMMPS package for classical molecular dynamics simulations. The Verlet velocity integrator with a timestep of 2~fs was used. The temperature and pressure were controlled using Nosé–Hoover thermostat with $\tau=0.1$~ps for temperature and $\tau=1$~ps for pressure. Every simulation was performed in 15000 timesteps.

\begin{figure} [h]
  \centering
  \includegraphics [width=0.99\columnwidth]{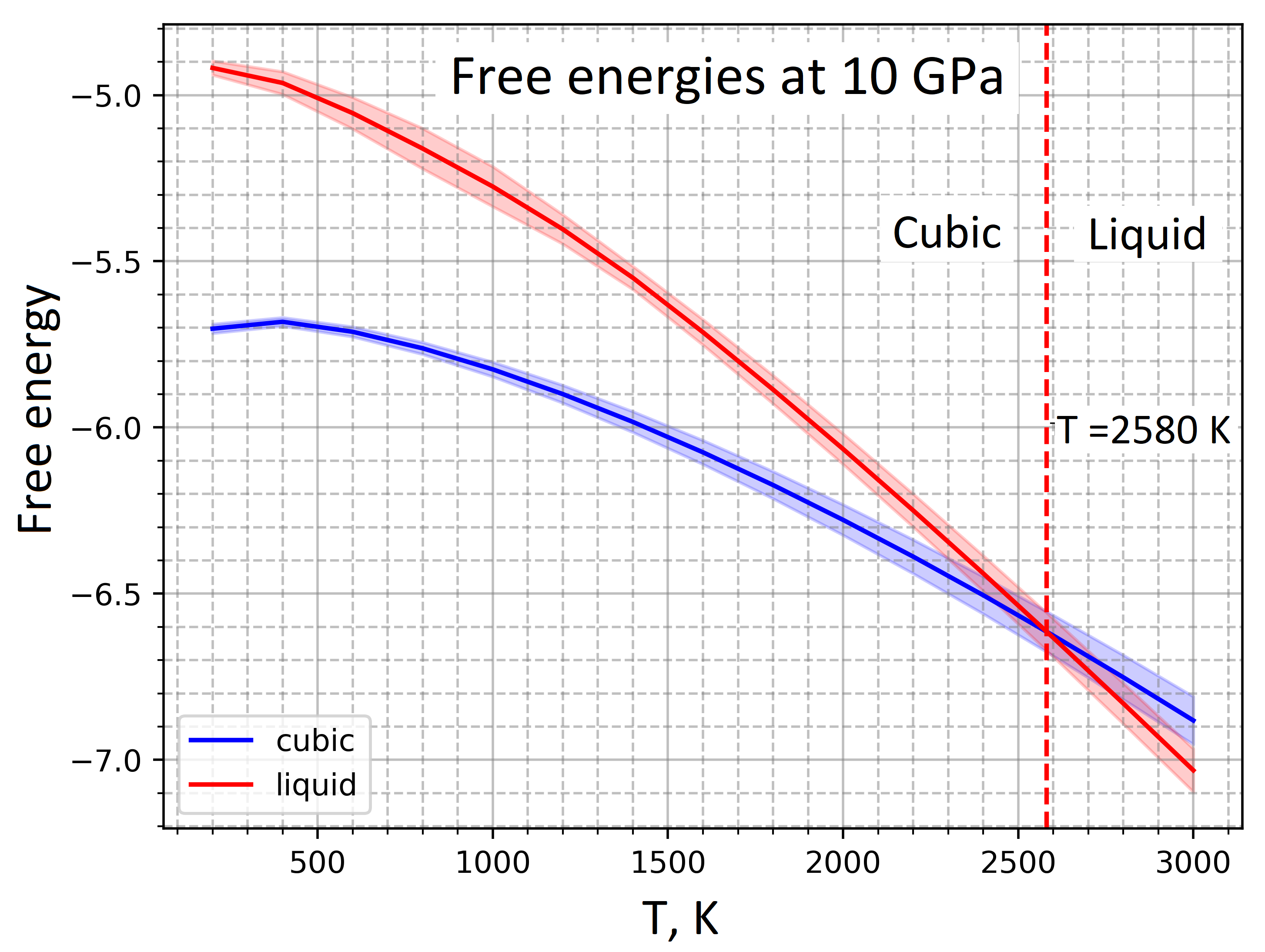}
  \caption{Initial point (2580~K, 10~GPa) used in Gibbs--Duhem integration to find the melting temperature in the pressure range from 0 to 50~GPa.}
  \label{fig:init-point}
\end{figure}

All further calculations were carried out using the Deep Potential Thermodynamic Integration (DPTI) software~\cite{dpti2020}. We performed Hamiltonian thermodynamic integration (HTI) where the Hamiltonian of the system changes between the initial ($\lambda = 0$) and the final ($\lambda = 1$) states. Then the free energy of the system becomes a function not only of the thermodynamic variables but also of the coupling parameter $\lambda$. The HTI was performed at 1000~K for zb-BP and at 3000~K for the liquid phase. The Gibbs free energies for both phases were found in the range from 100 to 3000~K by thermodynamic integration. Then Gibbs--Duhem integration (GDI) was applied, which allows the determination of the coexistence lines once an initial coexistence point is known. The intersection point located at 2580~K and 10~GPa (Fig.~\ref{fig:init-point}) was used as the GDI initial point to find the melting point in the range from 0 to 50~GPa. In this way, a melting curve (pressure dependence of the melting temperature, $T_\mathrm{m}(p)$) for BP was obtained.

All our DPTI calculations used the same thermostat and timestep as above. The HTI, TI and GDI performed 15000 timesteps every simulation. The Lennard--Jones potential parameters for TI were found using a radial distribution function for liquid BP to be $\sigma_{00}=1.3$~\AA, $\sigma_{01}=1.5$~\AA, and $\sigma_{11}=1.8$~\AA. All the $\lambda$ values used for HTI are listed in  Table~\ref{table-TI}.

In addition to the TI method, the developed DP potential was also applied to construct the melting curve by the method of two-phase molecular-dynamics simulations of coexisting solid and liquid~\cite{Zou2020,Rozas2022,Sun2022}. In this case, the melting point was determined by examining the movement of the solid--liquid interface during simulations. The results practically do not differ from those obtained by the TI method (see red stars in Fig.~\ref{fig:Tm}).

\subsection{Melting curve: Calculation and experiment}

\begin{figure} [h]
  \centering
  \includegraphics [width=0.95\columnwidth]{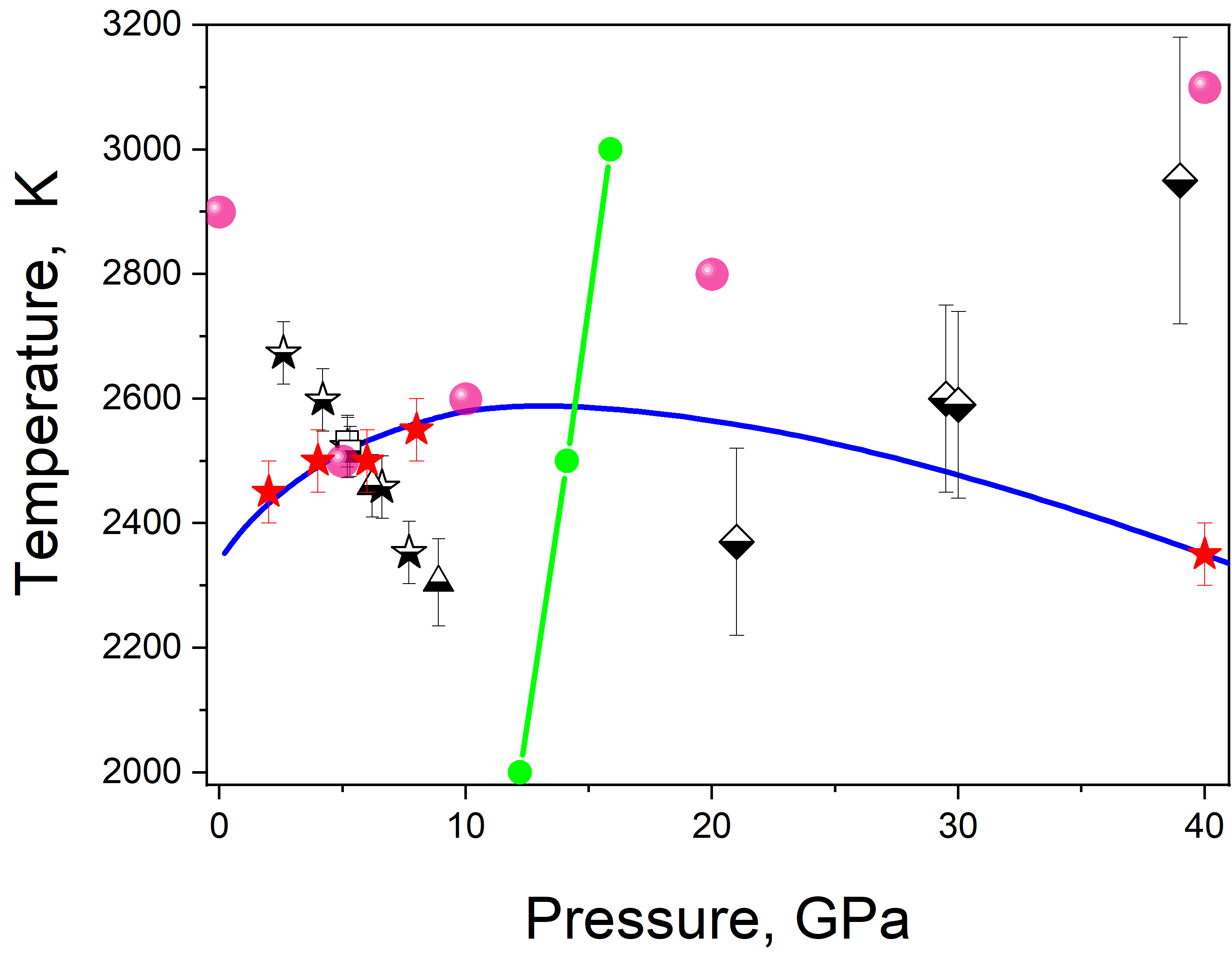}
  \caption{The melting curve of boron phosphide. Half-filled symbols are the experimental data: triangles, squares and stars taken from Ref.~\cite{Solozhenko2015}, diamonds – present work. Solid blue curve represents the results of DP-based thermodynamic integration calculations; red stars are the results of DP-based solid-liquid interface simulations. Pink spheres are the results estimation via AIMD simulations. Green bullets demonstrate approximate locus of structural and density anomalies in liquid BP.}
  \label{fig:Tm}
\end{figure}

Figure~\ref{fig:Tm} shows simulated and experimental pressure dependencies of the melting point of BP. Half-filled symbols denote experimental data obtained in~\cite{Solozhenko2015} (below 10~GPa) and in this work (above 20~GPa). The results of synchrotron X-ray diffraction experiments are represented by squares (HASYLAB-DESY), triangles (ESRF) and diamonds (SPring-8, present work). The stars indicate the onset of melting as registered \textit{in situ} by electrical resistivity measurements~\cite{Solozhenko2015}.

It can be seen that the experimental points start at $P=2.6$~GPa. The reason is that at ambient pressure, decomposition of boron phosphide is observed already at 1400~K, and the ambient-pressure melting point can not be experimentally determined. The error bars in the figure increase significantly at higher pressures, indicating the difficulty of experimentally studying the melting of a highly compressed BP in this region due to very high temperatures.

The technique for theoretically determining the melting curve $T_{\mathrm{m}}(P)$ is described above in Section~III. The results of our DP-based simulations (dashed red line) demonstrate a reverse trend in evolution of the melting point with pressure. Indeed, The simulated $T_{\mathrm{m}}(P)$ curve develop a maximum at $P \approx 12$ GPa. The location of this maximum correlate closely with the locus of both density and structural anomalies in the liquid state (blue stars).It is also important that the presence of density anomaly at the maximum of the melting curve is necessary to satisfy the Clausius–Clapeyron relation. In that sense, despite the disagreement with experimental data, our calculations are in agreement with the behaviour of structural and thermodynamical properties. The results of our AIMD simulations (pink spheres) in low and high ends of the pressure range considered are in qualitative agreement with experiment. However, this agreement is not due some technical issues (see discussion in sec.~\ref{sec:discussion}).


\begin{figure} [h]
  \centering
  \includegraphics [width=0.99\columnwidth]{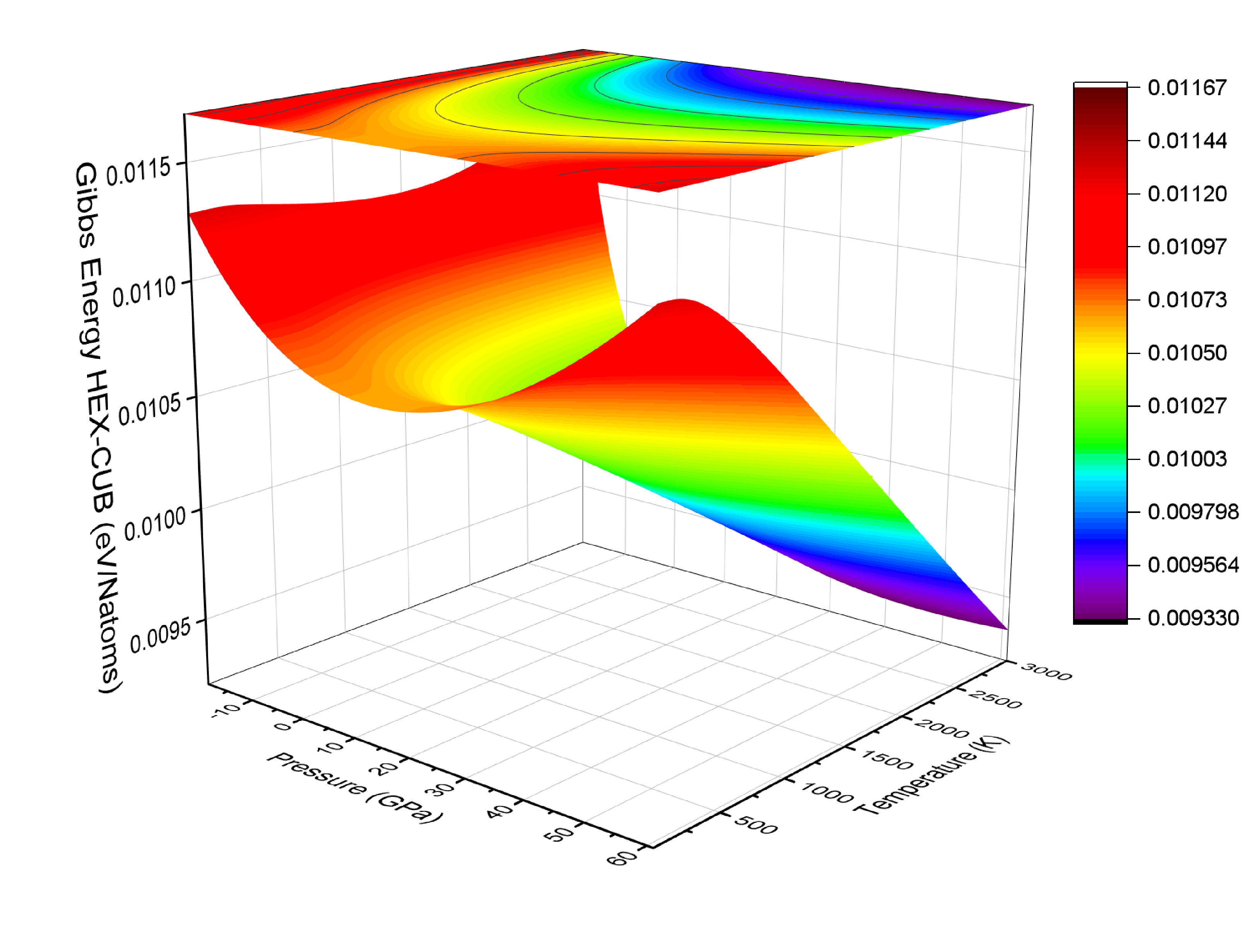}
  \caption{Gibbs energy difference between the hexagonal and cubic phases of BP, $G_{\mathrm{hex}}$--$G_{\mathrm{cub}}$.}
  \label{fig:gibbs}
\end{figure}

\subsection{On the transition between cubic and hexagonal BP phases}

In addition to the cubic phase zb-BP, we studied the hexagonal wurtzite (w-BP) in an attempt to determine its thermodynamic stability region. Recall that data on this phase in the literature are very scarce. Figure~\ref{fig:gibbs} shows the difference in the Gibbs free energy ($G$) between the two BP polymorphs. It turned out that the Gibbs energy of w-BP, $G_{\mathrm{hex}}$, exceeds $G_{\mathrm{cub}}$ over the entire, rather wide, $p$--$T$ region. Interestingly, at intermediate pressures on the order of 10--20~GPa, an anomaly is observed -- nonmonotonic behavior of the difference $G_{\mathrm{hex}}$-- $G_{\mathrm{cub}}$. Note that the absolute values of this difference are rather small and they are of the order of accuracy of Gibbs energy calculations in the framework of DP. These results open up issues regarding experimental stability of the hexagonal BP phase and its theoretical determination.



\section{Discussion and conclusions \label{sec:discussion}}
In this work, we address developing deep machine learning potentials for simulating covalent materials with nonisotropic interparticle interactions. We consider boron phosphide as an example of such systems, which is interesting from both fundamental and practical points of view. By using the DeePMD package as a constructor of neural network potentials, the VASP package as an \textit{ab initio} calculator, the LAMMPS package for classical molecular dynamics simulations and the DPGEN tool for realization an active learning strategy, we develop a deep machine learning potential (DP) which allows simulating the behavior of boron phosphide in liquid and solid states in the temperature range of 2000--4000~K and pressures of 0--40~GPa. The potential gives reasonable agreement with experiential and \textit{ab initio} molecular dynamics results for radial distribution functions, velocity autocorrelation functions and equations of states (see Figs.~\ref{fig:rdf}, \ref{fig:vaf}, \ref{fig:isoth-eos}).


Our DP-based simulations reveal that the structure of BP changes drastically during melting (see Fig.\ref{fig:liq_struct}).  Namely, a locally tetrahedral cubic BP crystal transforms into an open network which exhibits a strong tendency for clustering of the same species and actually consists of two interpenetrating boron and phosphorous sub-networks with different structures. One of the consequences of such structural changes is the drastic decrease in the density of BP (about 30\%).

This structure of liquid BP differs substantially from the structure of its constituents in liquid states. Indeed, the structure of liquid boron is characterized by six average coordination numbers with the pentagonal pyramidal polyhedrons as a primary building unit \cite{Price2009PRL,Durandurdu2015JNCS} whereas P melt at normal pressure consists of tetrahedral ${P_4}$ molecules \cite{Senda2002JPCM, Ghiringhelli2005JCP}. This suggests that B- and P-networks in BP melts cannot be considered separately as different independent objects but they are the result of complicated chemical interaction between these species. The observed structure of BP melt is different from that for many other tetrahedrally bonded semiconductors that become metallic and more closed-packed under melting~\cite{Porowski2019JCrystGrowth}. On the other hand, some systems, for example GaN, reveal similar behaviour~\cite{Porowski2019JCrystGrowth}.  This suggests a complicated nature of melts in materials with strongly directed covalent interactions which can be essentially system-dependent.



We find that the structure of the boron phosphide melts changes significantly as the pressure increases. In terms of coordination numbers, these changes are reflected by the evolution from low-pressure tetrahedral-like coordination ($Z\approx 4$) to high-pressure octahedral-like coordination ($Z \approx 6$). The analysis of pressure evolution of the local structure (RDFs, BADFs, coordination numbers, bond lengths) reveals that the main contributions to structural changes at low pressures are made by the evolution of the medium-range order in the B-subnetwork and at high pressures by the change of the short-range order in P-subnetwork.  Particularly, noticeable changes occur at intermediate pressures where the partial P-P coordination numbers and bond lengths exhibit an anomalous behavior in the range of 12--15~GPa. For a not very clear reason, an anomalous nonmonotonicity in the Gibbs energy difference between the hexagonal and cubic phases of BP, $G_{\mathrm{hex}}$--$G_{\mathrm{cub}}$, is also observed at 10--20~GPa.

Unexpected results are obtained for melting temperature as a function of pressure. DP-based simulations reveal that $T_m(P)$ curve develops a maximum at $P\approx 13$ GPa, whereas experimental studies provide two separate branches of the melting curve, which demonstrate an opposite behaviour (see Fig.~\ref{fig:Tm}). The reasons for this contradiction are not clear yet. One can suggest that the main problem lies in the field of MD simulations and is caused by some inaccuracies of interatomic potential. Indeed, calculating the melting temperatures is a challenging task for \textit{ab initio} simulations because their values are very sensitive to possible inaccuracies in determining the free energies and chemical potentials of the solid and liquid phases. This is a well-known problem. For example, for tin, the deviation of the DFT origin melting line~\cite{dpti2020} from the experimental one exceeds 100\%~\cite{Xu_2014}; better accuracy has been achieved while building MLIP via SCAN meta-GGA functionals rather than GGA ones~\cite{dpti2020}. Accurate calculation of the melting point of Si requires going even further -- to RPA \textit{ab initio} calculations~\cite{Dorner2018}. Considering these cases, the melting line obtained by DP built via GGA DFT should be considered quite good. Better accuracy is expected when building DP via either SCAN meta-GGA or hybrid functionals; this is likely to be done in the future.

The melting curve was also estimated by AIMD (without using DP) through direct simulations of melting in overheated crystals. The results obtained seem to be in qualitative  agreement with experimental data and differ substantially from the results of DP-based simulations (see pink bullets in Fig.~\ref{fig:Tm}). However, it should be noted, that such a method of $T_m$ determinations is highly inaccurate due to a strong tendency of overheating in simulations of small crystal supercells and an impossibility to collect many independent statistical data. We present these data only to demonstrate problems of simulations of phase transitions via AIMD and to underline the importance of machine learning-based simulations.

The above calculation problems are expected to affect absolute values of the melting temperatures but we believe that DP simulations provide correct qualitative behaviour of $T_m(P)$ (in particular the maximum). This point is supported by the fact that the behaviour of the melting line is in agreement with structural changes in melts and the equations of states (and so the Clausius–Clapeyron relation). If so, the question arises: what is the reason for the discrepancy between the experiment and DP simulations? A possible answer can be found in the discussion recently reported in Ref.~\cite{Porowski2019JCrystGrowth} where similar behaviour of the melting line was observed for GaN. The authors analyse the melting behaviour of different tetrahedrally bonded semiconductors and suggest that the maximum on a $T_m(P)$ curve is probably a universal phenomenon for such systems. The fact that this phenomenon is rarely observed is due to a possible shift of this maximum in the range of negative pressures. Our main hypothesis is that the same is true for BP. Thus, the calculated $T_m(P)$ curve is qualitatively true but its maximum, which should be at negative pressures, is shifted in the range of positive pressures due to inaccuracies in $T_m$ calculations. These hypotheses stimulate further experimental and theoretical  studies: it is interesting to perform experimental studies and negative pressures and develop a more accurate DP potential (for example using meta-GGA calculations) to quantitatively calculate  $T_m(P)$.


It can be concluded that our results show the promise of using deep learning potentials for simulating covalent materials stimulate further studies of the properties of boron phosphide at high pressures including negative ones.

\section {Acknowledgments}
V.L.S. and K.A.C. are thankful to Dr.~Saori~I.~Kawaguchi for assistance in laser-heated DAC experiments at BL10XU beamline. This work was supported by the Russian Science Foundation under Grant RSF \#22-22-00806 (https://rscf.ru/en/project/22-22-00806/). Synchrotron X-ray diffraction studies have been performed during beamtime allocated to proposal 2019A1054 at SPring-8. Numerical calculations were performed using computing resources of the federal collective usage center Complex for Simulation and Data Processing for Mega-science Facilities at NRC ”Kurchatov Institute” (http://ckp.nrcki.ru/), supercomputers at Joint Supercomputer Center of RAS (JSCC RAS) and the 'Uran' cluster of IMM UB RAS (https://parallel.uran.ru/).
\bibliography{bp-refs}  

\begin{thebibliography}{56}%
\makeatletter
\providecommand \@ifxundefined [1]{%
 \@ifx{#1\undefined}
}%
\providecommand \@ifnum [1]{%
 \ifnum #1\expandafter \@firstoftwo
 \else \expandafter \@secondoftwo
 \fi
}%
\providecommand \@ifx [1]{%
 \ifx #1\expandafter \@firstoftwo
 \else \expandafter \@secondoftwo
 \fi
}%
\providecommand \natexlab [1]{#1}%
\providecommand \enquote  [1]{``#1''}%
\providecommand \bibnamefont  [1]{#1}%
\providecommand \bibfnamefont [1]{#1}%
\providecommand \citenamefont [1]{#1}%
\providecommand \href@noop [0]{\@secondoftwo}%
\providecommand \href [0]{\begingroup \@sanitize@url \@href}%
\providecommand \@href[1]{\@@startlink{#1}\@@href}%
\providecommand \@@href[1]{\endgroup#1\@@endlink}%
\providecommand \@sanitize@url [0]{\catcode `\\12\catcode `\$12\catcode
  `\&12\catcode `\#12\catcode `\^12\catcode `\_12\catcode `\%12\relax}%
\providecommand \@@startlink[1]{}%
\providecommand \@@endlink[0]{}%
\providecommand \url  [0]{\begingroup\@sanitize@url \@url }%
\providecommand \@url [1]{\endgroup\@href {#1}{\urlprefix }}%
\providecommand \urlprefix  [0]{URL }%
\providecommand \Eprint [0]{\href }%
\providecommand \doibase [0]{https://doi.org/}%
\providecommand \selectlanguage [0]{\@gobble}%
\providecommand \bibinfo  [0]{\@secondoftwo}%
\providecommand \bibfield  [0]{\@secondoftwo}%
\providecommand \translation [1]{[#1]}%
\providecommand \BibitemOpen [0]{}%
\providecommand \bibitemStop [0]{}%
\providecommand \bibitemNoStop [0]{.\EOS\space}%
\providecommand \EOS [0]{\spacefactor3000\relax}%
\providecommand \BibitemShut  [1]{\csname bibitem#1\endcsname}%
\let\auto@bib@innerbib\@empty
\bibitem [{\citenamefont {Solozhenko}\ and\ \citenamefont
  {Mukhanov}(2015)}]{Solozhenko2015}%
  \BibitemOpen
  \bibfield  {author} {\bibinfo {author} {\bibfnamefont {V.~L.}\ \bibnamefont
  {Solozhenko}}\ and\ \bibinfo {author} {\bibfnamefont {V.~A.}\ \bibnamefont
  {Mukhanov}},\ }\bibfield  {title} {\bibinfo {title} {On melting of boron
  phosphide under pressure},\ }\href
  {https://doi.org/10.3103/S1063457615060106} {\bibfield  {journal} {\bibinfo
  {journal} {J. Superhard Mater.}\ }\textbf {\bibinfo {volume} {37}},\ \bibinfo
  {pages} {438} (\bibinfo {year} {2015})}\BibitemShut {NoStop}%
\bibitem [{\citenamefont {Behler}(2016)}]{Behler2016}%
  \BibitemOpen
  \bibfield  {author} {\bibinfo {author} {\bibfnamefont {J.}~\bibnamefont
  {Behler}},\ }\bibfield  {title} {\bibinfo {title} {Perspective: Machine
  learning potentials for atomistic simulations},\ }\href
  {https://doi.org/10.1063/1.4966192} {\bibfield  {journal} {\bibinfo
  {journal} {J. Chem. Phys.}\ }\textbf {\bibinfo {volume} {145}},\ \bibinfo
  {pages} {170901} (\bibinfo {year} {2016})}\BibitemShut {NoStop}%
\bibitem [{\citenamefont {Jinnouchi}\ \emph {et~al.}()\citenamefont
  {Jinnouchi}, \citenamefont {Karsai},\ and\ \citenamefont
  {Kresse}}]{Jinnouchi}%
  \BibitemOpen
  \bibfield  {author} {\bibinfo {author} {\bibfnamefont {R.}~\bibnamefont
  {Jinnouchi}}, \bibinfo {author} {\bibfnamefont {F.}~\bibnamefont {Karsai}},\
  and\ \bibinfo {author} {\bibfnamefont {G.}~\bibnamefont {Kresse}},\
  }\href@noop {} {\bibfield  {journal} {\bibinfo  {journal} {Phys. Rev. B}\
  }\textbf {\bibinfo {volume} {100}},\ \bibinfo {pages} {14105}}\BibitemShut
  {NoStop}%
\bibitem [{\citenamefont {Behler}\ and\ \citenamefont
  {Csányi}(2021)}]{Behler2021}%
  \BibitemOpen
  \bibfield  {author} {\bibinfo {author} {\bibfnamefont {J.}~\bibnamefont
  {Behler}}\ and\ \bibinfo {author} {\bibfnamefont {G.}~\bibnamefont
  {Csányi}},\ }\bibfield  {title} {\bibinfo {title} {Machine learning
  potentials for extended systems: a perspective},\ }\href
  {https://doi.org/10.1140/epjb/s10051-021-00156-1} {\bibfield  {journal}
  {\bibinfo  {journal} {Eur. Phys. J. B}\ }\textbf {\bibinfo {volume} {94}},\
  \bibinfo {pages} {142} (\bibinfo {year} {2021})}\BibitemShut {NoStop}%
\bibitem [{\citenamefont {Vandermause}\ \emph {et~al.}(2020)\citenamefont
  {Vandermause}, \citenamefont {Torrisi}, \citenamefont {Batzner},
  \citenamefont {Xie}, \citenamefont {Sun}, \citenamefont {Kolpak},\ and\
  \citenamefont {Kozinsky}}]{Vandermause2020}%
  \BibitemOpen
  \bibfield  {author} {\bibinfo {author} {\bibfnamefont {J.}~\bibnamefont
  {Vandermause}}, \bibinfo {author} {\bibfnamefont {S.}~\bibnamefont
  {Torrisi}}, \bibinfo {author} {\bibfnamefont {S.}~\bibnamefont {Batzner}},
  \bibinfo {author} {\bibfnamefont {Y.}~\bibnamefont {Xie}}, \bibinfo {author}
  {\bibfnamefont {L.}~\bibnamefont {Sun}}, \bibinfo {author} {\bibfnamefont
  {A.}~\bibnamefont {Kolpak}},\ and\ \bibinfo {author} {\bibfnamefont
  {B.}~\bibnamefont {Kozinsky}},\ }\bibfield  {title} {\bibinfo {title}
  {On-the-fly active learning of interpretable bayesian force fields for
  atomistic rare events},\ }\bibfield  {journal} {\bibinfo  {journal} {npj
  Comput. Mater.}\ }\textbf {\bibinfo {volume} {6}},\ \href
  {https://doi.org/10.1038/s41524-020-0283-z} {10.1038/s41524-020-0283-z}
  (\bibinfo {year} {2020})\BibitemShut {NoStop}%
\bibitem [{\citenamefont {Choudhary}\ and\ \citenamefont
  {Decost}(2021)}]{Choudhary2021}%
  \BibitemOpen
  \bibfield  {author} {\bibinfo {author} {\bibfnamefont {K.}~\bibnamefont
  {Choudhary}}\ and\ \bibinfo {author} {\bibfnamefont {B.}~\bibnamefont
  {Decost}},\ }\bibfield  {title} {\bibinfo {title} {Atomistic line graph
  neural network for improved materials property predictions},\ }\bibfield
  {journal} {\bibinfo  {journal} {npj Comput. Mater.}\ }\textbf {\bibinfo
  {volume} {7}},\ \href {https://doi.org/10.1038/s41524-021-00650-1}
  {10.1038/s41524-021-00650-1} (\bibinfo {year} {2021})\BibitemShut {NoStop}%
\bibitem [{\citenamefont {Choudhary}\ \emph {et~al.}(2022)\citenamefont
  {Choudhary}, \citenamefont {Decost}, \citenamefont {Chen}, \citenamefont
  {Jain}, \citenamefont {Tavazza}, \citenamefont {Cohn}, \citenamefont {Park},
  \citenamefont {Choudhary}, \citenamefont {Agrawal}, \citenamefont {Billinge},
  \citenamefont {Holm}, \citenamefont {Ong},\ and\ \citenamefont
  {Wolverton}}]{Choudhary2022}%
  \BibitemOpen
  \bibfield  {author} {\bibinfo {author} {\bibfnamefont {K.}~\bibnamefont
  {Choudhary}}, \bibinfo {author} {\bibfnamefont {B.}~\bibnamefont {Decost}},
  \bibinfo {author} {\bibfnamefont {C.}~\bibnamefont {Chen}}, \bibinfo {author}
  {\bibfnamefont {A.}~\bibnamefont {Jain}}, \bibinfo {author} {\bibfnamefont
  {F.}~\bibnamefont {Tavazza}}, \bibinfo {author} {\bibfnamefont
  {R.}~\bibnamefont {Cohn}}, \bibinfo {author} {\bibfnamefont {C.}~\bibnamefont
  {Park}}, \bibinfo {author} {\bibfnamefont {A.}~\bibnamefont {Choudhary}},
  \bibinfo {author} {\bibfnamefont {A.}~\bibnamefont {Agrawal}}, \bibinfo
  {author} {\bibfnamefont {S.}~\bibnamefont {Billinge}}, \bibinfo {author}
  {\bibfnamefont {E.}~\bibnamefont {Holm}}, \bibinfo {author} {\bibfnamefont
  {S.}~\bibnamefont {Ong}},\ and\ \bibinfo {author} {\bibfnamefont
  {C.}~\bibnamefont {Wolverton}},\ }\bibfield  {title} {\bibinfo {title}
  {Recent advances and applications of deep learning methods in materials
  science},\ }\bibfield  {journal} {\bibinfo  {journal} {npj Comput. Mater.}\
  }\textbf {\bibinfo {volume} {8}},\ \href
  {https://doi.org/10.1038/s41524-022-00734-6} {10.1038/s41524-022-00734-6}
  (\bibinfo {year} {2022})\BibitemShut {NoStop}%
\bibitem [{\citenamefont {Minakov}\ \emph {et~al.}()\citenamefont {Minakov},
  \citenamefont {Paramonov}, \citenamefont {Demyanov}, \citenamefont {Fokin},\
  and\ \citenamefont {Levashov}}]{Minakov}%
  \BibitemOpen
  \bibfield  {author} {\bibinfo {author} {\bibfnamefont {D.}~\bibnamefont
  {Minakov}}, \bibinfo {author} {\bibfnamefont {M.}~\bibnamefont {Paramonov}},
  \bibinfo {author} {\bibfnamefont {G.}~\bibnamefont {Demyanov}}, \bibinfo
  {author} {\bibfnamefont {V.}~\bibnamefont {Fokin}},\ and\ \bibinfo {author}
  {\bibfnamefont {P.}~\bibnamefont {Levashov}},\ }\href@noop {} {\bibfield
  {journal} {\bibinfo  {journal} {Phys. Rev. B}\ }\textbf {\bibinfo {volume}
  {106}},\ \bibinfo {pages} {214105}}\BibitemShut {NoStop}%
\bibitem [{\citenamefont {Deshchenya}\ \emph {et~al.}(2022)\citenamefont
  {Deshchenya}, \citenamefont {Kondratyuk}, \citenamefont {Lankin},\ and\
  \citenamefont {Norman}}]{Deshchenya2022}%
  \BibitemOpen
  \bibfield  {author} {\bibinfo {author} {\bibfnamefont {V.}~\bibnamefont
  {Deshchenya}}, \bibinfo {author} {\bibfnamefont {N.}~\bibnamefont
  {Kondratyuk}}, \bibinfo {author} {\bibfnamefont {A.}~\bibnamefont {Lankin}},\
  and\ \bibinfo {author} {\bibfnamefont {G.}~\bibnamefont {Norman}},\
  }\bibfield  {title} {\bibinfo {title} {Molecular dynamics study of sucrose
  aqueous solutions: From solution structure to transport coefficients},\
  }\href {https://doi.org/10.1016/j.molliq.2022.120456} {\bibfield  {journal}
  {\bibinfo  {journal} {J. Mol. Liq.}\ }\textbf {\bibinfo {volume} {367}},\
  \bibinfo {pages} {120456} (\bibinfo {year} {2022})}\BibitemShut {NoStop}%
\bibitem [{\citenamefont {Antropov}\ and\ \citenamefont
  {Stegailov}(2020)}]{Antropov_2020}%
  \BibitemOpen
  \bibfield  {author} {\bibinfo {author} {\bibfnamefont {A.}~\bibnamefont
  {Antropov}}\ and\ \bibinfo {author} {\bibfnamefont {V.}~\bibnamefont
  {Stegailov}},\ }\bibfield  {title} {\bibinfo {title} {Nanobubbles diffusion
  in bcc uranium: Theory and atomistic modelling},\ }\href
  {https://doi.org/10.1016/j.jnucmat.2020.152110} {\bibfield  {journal}
  {\bibinfo  {journal} {J. Nucl. Mater.}\ }\textbf {\bibinfo {volume} {533}},\
  \bibinfo {pages} {152110} (\bibinfo {year} {2020})}\BibitemShut {NoStop}%
\bibitem [{\citenamefont {Behler}(2011)}]{Behler2011}%
  \BibitemOpen
  \bibfield  {author} {\bibinfo {author} {\bibfnamefont {J.}~\bibnamefont
  {Behler}},\ }\bibfield  {title} {\bibinfo {title} {Neural network
  potential-energy surfaces in chemistry: a tool for large-scale simulations},\
  }\href {https://doi.org/10.1039/C1CP21668F} {\bibfield  {journal} {\bibinfo
  {journal} {Phys. Chem. Chem. Phys.}\ }\textbf {\bibinfo {volume} {13}},\
  \bibinfo {pages} {17930} (\bibinfo {year} {2011})}\BibitemShut {NoStop}%
\bibitem [{\citenamefont {Behler}(2017)}]{Behler2017}%
  \BibitemOpen
  \bibfield  {author} {\bibinfo {author} {\bibfnamefont {J.}~\bibnamefont
  {Behler}},\ }\bibfield  {title} {\bibinfo {title} {First principles neural
  network potentials for reactive simulations of large molecular and condensed
  systems},\ }\href {https://doi.org/https://doi.org/10.1002/anie.201703114}
  {\bibfield  {journal} {\bibinfo  {journal} {Angew. Chem. Int. Ed.}\ }\textbf
  {\bibinfo {volume} {56}},\ \bibinfo {pages} {12828} (\bibinfo {year}
  {2017})}\BibitemShut {NoStop}%
\bibitem [{\citenamefont {Wang}\ \emph
  {et~al.}(2018{\natexlab{a}})\citenamefont {Wang}, \citenamefont {Zhang},
  \citenamefont {Han},\ and\ \citenamefont {Weinan}}]{Wang2018a}%
  \BibitemOpen
  \bibfield  {author} {\bibinfo {author} {\bibfnamefont {H.}~\bibnamefont
  {Wang}}, \bibinfo {author} {\bibnamefont {Zhang}}, \bibinfo {author}
  {\bibnamefont {Han}},\ and\ \bibinfo {author} {\bibnamefont {Weinan}},\
  }\bibfield  {title} {\bibinfo {title} {{DeePMD}-kit: A deep learning package
  for many-body potential energy representation and molecular dynamics},\
  }\href {https://doi.org/https://doi.org/10.1016/j.cpc.2018.03.016} {\bibfield
   {journal} {\bibinfo  {journal} {Comput. Phys. Commun.}\ }\textbf {\bibinfo
  {volume} {228}},\ \bibinfo {pages} {178} (\bibinfo {year}
  {2018}{\natexlab{a}})}\BibitemShut {NoStop}%
\bibitem [{\citenamefont {Deringer}\ \emph {et~al.}(2019)\citenamefont
  {Deringer}, \citenamefont {Caro},\ and\ \citenamefont
  {Csányi}}]{Deringer2019}%
  \BibitemOpen
  \bibfield  {author} {\bibinfo {author} {\bibfnamefont {V.~L.}\ \bibnamefont
  {Deringer}}, \bibinfo {author} {\bibfnamefont {M.~A.}\ \bibnamefont {Caro}},\
  and\ \bibinfo {author} {\bibfnamefont {G.}~\bibnamefont {Csányi}},\
  }\bibfield  {title} {\bibinfo {title} {Machine learning interatomic
  potentials as emerging tools for materials science},\ }\href
  {https://doi.org/https://doi.org/10.1002/adma.201902765} {\bibfield
  {journal} {\bibinfo  {journal} {Adv. Mater.}\ }\textbf {\bibinfo {volume}
  {31}},\ \bibinfo {pages} {1902765} (\bibinfo {year} {2019})}\BibitemShut
  {NoStop}%
\bibitem [{\citenamefont {Mishin}(2021)}]{Mishin2021}%
  \BibitemOpen
  \bibfield  {author} {\bibinfo {author} {\bibfnamefont {Y.}~\bibnamefont
  {Mishin}},\ }\bibfield  {title} {\bibinfo {title} {Machine-learning
  interatomic potentials for materials science},\ }\href
  {https://doi.org/https://doi.org/10.1016/j.actamat.2021.116980} {\bibfield
  {journal} {\bibinfo  {journal} {Acta Mater.}\ }\textbf {\bibinfo {volume}
  {214}},\ \bibinfo {pages} {116980} (\bibinfo {year} {2021})}\BibitemShut
  {NoStop}%
\bibitem [{\citenamefont {Duff}\ \emph {et~al.}(2015)\citenamefont {Duff},
  \citenamefont {Finnis}, \citenamefont {Maugis}, \citenamefont {Thijsse},\
  and\ \citenamefont {Sluiter}}]{Duff2015}%
  \BibitemOpen
  \bibfield  {author} {\bibinfo {author} {\bibfnamefont {A.}~\bibnamefont
  {Duff}}, \bibinfo {author} {\bibfnamefont {M.}~\bibnamefont {Finnis}},
  \bibinfo {author} {\bibfnamefont {P.}~\bibnamefont {Maugis}}, \bibinfo
  {author} {\bibfnamefont {B.}~\bibnamefont {Thijsse}},\ and\ \bibinfo {author}
  {\bibfnamefont {M.}~\bibnamefont {Sluiter}},\ }\bibfield  {title} {\bibinfo
  {title} {Meamfit: A reference-free modified embedded atom method (rf-meam)
  energy and force-fitting code},\ }\href
  {https://doi.org/10.1016/j.cpc.2015.05.016} {\bibfield  {journal} {\bibinfo
  {journal} {Comput. Phys. Commun.}\ }\textbf {\bibinfo {volume} {196}},\
  \bibinfo {pages} {439} (\bibinfo {year} {2015})}\BibitemShut {NoStop}%
\bibitem [{\citenamefont {Balyakin}\ \emph {et~al.}(2020)\citenamefont
  {Balyakin}, \citenamefont {Rempel}, \citenamefont {Ryltsev},\ and\
  \citenamefont {Rempel}}]{Balyakin2020}%
  \BibitemOpen
  \bibfield  {author} {\bibinfo {author} {\bibfnamefont {I.~A.}\ \bibnamefont
  {Balyakin}}, \bibinfo {author} {\bibfnamefont {S.~V.}\ \bibnamefont
  {Rempel}}, \bibinfo {author} {\bibfnamefont {R.~E.}\ \bibnamefont
  {Ryltsev}},\ and\ \bibinfo {author} {\bibfnamefont {A.~A.}\ \bibnamefont
  {Rempel}},\ }\bibfield  {title} {\bibinfo {title} {Deep machine learning
  interatomic potential for liquid silica},\ }\href
  {https://doi.org/10.1103/PhysRevE.102.052125} {\bibfield  {journal} {\bibinfo
   {journal} {Phys. Rev. E}\ }\textbf {\bibinfo {volume} {102}},\ \bibinfo
  {pages} {052125} (\bibinfo {year} {2020})}\BibitemShut {NoStop}%
\bibitem [{\citenamefont {Wang}\ \emph {et~al.}(2021)\citenamefont {Wang},
  \citenamefont {Wang}, \citenamefont {Zhao}, \citenamefont {Du}, \citenamefont
  {Xu}, \citenamefont {Gu},\ and\ \citenamefont {Duan}}]{Wang2021}%
  \BibitemOpen
  \bibfield  {author} {\bibinfo {author} {\bibfnamefont {Z.}~\bibnamefont
  {Wang}}, \bibinfo {author} {\bibfnamefont {C.}~\bibnamefont {Wang}}, \bibinfo
  {author} {\bibfnamefont {S.}~\bibnamefont {Zhao}}, \bibinfo {author}
  {\bibfnamefont {S.}~\bibnamefont {Du}}, \bibinfo {author} {\bibfnamefont
  {Y.}~\bibnamefont {Xu}}, \bibinfo {author} {\bibfnamefont {B.-L.}\
  \bibnamefont {Gu}},\ and\ \bibinfo {author} {\bibfnamefont {W.}~\bibnamefont
  {Duan}},\ }\bibfield  {title} {\bibinfo {title} {Symmetry-adapted graph
  neural networks for constructing molecular dynamics force fields},\
  }\bibfield  {journal} {\bibinfo  {journal} {Science China Physics, Mechanics
  and Astronomy}\ }\textbf {\bibinfo {volume} {64}},\ \href
  {https://doi.org/10.1007/s11433-021-1739-4} {10.1007/s11433-021-1739-4}
  (\bibinfo {year} {2021})\BibitemShut {NoStop}%
\bibitem [{\citenamefont {Malosso}\ \emph {et~al.}(2022)\citenamefont
  {Malosso}, \citenamefont {Zhang}, \citenamefont {Car}, \citenamefont
  {Baroni},\ and\ \citenamefont {Tisi}}]{Malosso_2022}%
  \BibitemOpen
  \bibfield  {author} {\bibinfo {author} {\bibfnamefont {C.}~\bibnamefont
  {Malosso}}, \bibinfo {author} {\bibfnamefont {L.}~\bibnamefont {Zhang}},
  \bibinfo {author} {\bibfnamefont {R.}~\bibnamefont {Car}}, \bibinfo {author}
  {\bibfnamefont {S.}~\bibnamefont {Baroni}},\ and\ \bibinfo {author}
  {\bibfnamefont {D.}~\bibnamefont {Tisi}},\ }\bibfield  {title} {\bibinfo
  {title} {Viscosity in water from first-principles and deep-neural-network
  simulations},\ }\bibfield  {booktitle} {\emph {\bibinfo {booktitle}
  {Viscosity in water from first-principles and deep-neural-network
  simulations}},\ }\bibfield  {journal} {\bibinfo  {journal} {npj Comput.
  Mater.}\ }\textbf {\bibinfo {volume} {8}},\ \href
  {https://doi.org/https://doi.org/10.1038/s41524-022-00830-7}
  {https://doi.org/10.1038/s41524-022-00830-7} (\bibinfo {year}
  {2022})\BibitemShut {NoStop}%
\bibitem [{\citenamefont {Villars}\ and\ \citenamefont
  {Calvert}()}]{Villars1985}%
  \BibitemOpen
  \bibinfo {editor} {\bibfnamefont {P.}~\bibnamefont {Villars}}\ and\ \bibinfo
  {editor} {\bibfnamefont {L.}~\bibnamefont {Calvert}},\ eds.,\ \href@noop {}
  {\emph {\bibinfo {title} {Pearson's Handbook of Crystallographic Data for
  Intermetallic Phases}}}\BibitemShut {NoStop}%
\bibitem [{\citenamefont {Stone}\ and\ \citenamefont {Hill}(1960)}]{Stone1960}%
  \BibitemOpen
  \bibfield  {author} {\bibinfo {author} {\bibfnamefont {B.}~\bibnamefont
  {Stone}}\ and\ \bibinfo {author} {\bibfnamefont {D.}~\bibnamefont {Hill}},\
  }\bibfield  {title} {\bibinfo {title} {Semiconducting properties of cubic
  boron phosphide},\ }\href {https://doi.org/10.1103/PhysRevLett.4.282}
  {\bibfield  {journal} {\bibinfo  {journal} {Phys. Rev. Lett.}\ }\textbf
  {\bibinfo {volume} {4}},\ \bibinfo {pages} {282} (\bibinfo {year}
  {1960})}\BibitemShut {NoStop}%
\bibitem [{\citenamefont {Solozhenko}\ and\ \citenamefont
  {Bushlya}(2019)}]{Bushlya2019}%
  \BibitemOpen
  \bibfield  {author} {\bibinfo {author} {\bibfnamefont {V.~L.}\ \bibnamefont
  {Solozhenko}}\ and\ \bibinfo {author} {\bibfnamefont {V.}~\bibnamefont
  {Bushlya}},\ }\bibfield  {title} {\bibinfo {title} {Mechanical properties of
  boron phosphides},\ }\href {https://doi.org/10.3103/S1063457619020023}
  {\bibfield  {journal} {\bibinfo  {journal} {J. Superhard Mater.}\ }\textbf
  {\bibinfo {volume} {41}},\ \bibinfo {pages} {84} (\bibinfo {year}
  {2019})}\BibitemShut {NoStop}%
\bibitem [{\citenamefont {Solozhenko}\ \emph {et~al.}(2014)\citenamefont
  {Solozhenko}, \citenamefont {Kurakevych}, \citenamefont {Le~Godec},
  \citenamefont {Kurnosov},\ and\ \citenamefont {Oganov}}]{Oganov2014}%
  \BibitemOpen
  \bibfield  {author} {\bibinfo {author} {\bibfnamefont {V.~L.}\ \bibnamefont
  {Solozhenko}}, \bibinfo {author} {\bibfnamefont {O.~O.}\ \bibnamefont
  {Kurakevych}}, \bibinfo {author} {\bibfnamefont {Y.}~\bibnamefont
  {Le~Godec}}, \bibinfo {author} {\bibfnamefont {A.~V.}\ \bibnamefont
  {Kurnosov}},\ and\ \bibinfo {author} {\bibfnamefont {A.~R.}\ \bibnamefont
  {Oganov}},\ }\bibfield  {title} {\bibinfo {title} {Boron phosphide under
  pressure: In situ study by raman scattering and x-ray diffraction},\ }\href
  {https://doi.org/10.1063/1.4890231} {\bibfield  {journal} {\bibinfo
  {journal} {J. Appl. Phys.}\ }\textbf {\bibinfo {volume} {116}},\ \bibinfo
  {pages} {033501} (\bibinfo {year} {2014})}\BibitemShut {NoStop}%
\bibitem [{\citenamefont {Kumashiro}\ \emph {et~al.}(1989)\citenamefont
  {Kumashiro}, \citenamefont {Mitsuhashi}, \citenamefont {Okaya}, \citenamefont
  {Muta}, \citenamefont {Koshiro}, \citenamefont {Takahashi},\ and\
  \citenamefont {Mirabayashi}}]{Kumashiro1989}%
  \BibitemOpen
  \bibfield  {author} {\bibinfo {author} {\bibfnamefont {Y.}~\bibnamefont
  {Kumashiro}}, \bibinfo {author} {\bibfnamefont {T.}~\bibnamefont
  {Mitsuhashi}}, \bibinfo {author} {\bibfnamefont {S.}~\bibnamefont {Okaya}},
  \bibinfo {author} {\bibfnamefont {F.}~\bibnamefont {Muta}}, \bibinfo {author}
  {\bibfnamefont {T.}~\bibnamefont {Koshiro}}, \bibinfo {author} {\bibfnamefont
  {Y.}~\bibnamefont {Takahashi}},\ and\ \bibinfo {author} {\bibfnamefont
  {M.}~\bibnamefont {Mirabayashi}},\ }\bibfield  {title} {\bibinfo {title}
  {Thermal conductivity of a boron phosphide single‐crystal wafer up to high
  temperature},\ }\href {https://doi.org/10.1063/1.342867} {\bibfield
  {journal} {\bibinfo  {journal} {J. Appl. Phys.}\ }\textbf {\bibinfo {volume}
  {65}},\ \bibinfo {pages} {2147} (\bibinfo {year} {1989})}\BibitemShut
  {NoStop}%
\bibitem [{\citenamefont {Yugo}\ and\ \citenamefont {Kimura}(1980)}]{Yugo1980}%
  \BibitemOpen
  \bibfield  {author} {\bibinfo {author} {\bibfnamefont {S.}~\bibnamefont
  {Yugo}}\ and\ \bibinfo {author} {\bibfnamefont {T.}~\bibnamefont {Kimura}},\
  }\bibfield  {title} {\bibinfo {title} {Thermoelectric power of boron
  phosphide at high temperatures},\ }\href
  {https://doi.org/https://doi.org/10.1002/pssa.2210590148} {\bibfield
  {journal} {\bibinfo  {journal} {Phys. Status Solidi A}\ }\textbf {\bibinfo
  {volume} {59}},\ \bibinfo {pages} {363} (\bibinfo {year} {1980})}\BibitemShut
  {NoStop}%
\bibitem [{\citenamefont {Woo}\ \emph {et~al.}(2016)\citenamefont {Woo},
  \citenamefont {Lee},\ and\ \citenamefont {Kovnir}}]{Woo2016}%
  \BibitemOpen
  \bibfield  {author} {\bibinfo {author} {\bibfnamefont {K.}~\bibnamefont
  {Woo}}, \bibinfo {author} {\bibfnamefont {K.}~\bibnamefont {Lee}},\ and\
  \bibinfo {author} {\bibfnamefont {K.}~\bibnamefont {Kovnir}},\ }\bibfield
  {title} {\bibinfo {title} {Bp: synthesis and properties of boron phosphide},\
  }\href {https://doi.org/10.1088/2053-1591/3/7/074003} {\bibfield  {journal}
  {\bibinfo  {journal} {Mater. Res. Express}\ }\textbf {\bibinfo {volume}
  {3}},\ \bibinfo {pages} {074003} (\bibinfo {year} {2016})}\BibitemShut
  {NoStop}%
\bibitem [{\citenamefont {Gui}\ \emph {et~al.}(2020)\citenamefont {Gui},
  \citenamefont {Xue}, \citenamefont {Zhou}, \citenamefont {Gu}, \citenamefont
  {Ren}, \citenamefont {Cheng}, \citenamefont {Ma}, \citenamefont {Qin},
  \citenamefont {Liang}, \citenamefont {Yan}, \citenamefont {Zhang},
  \citenamefont {Zhang}, \citenamefont {Yu}, \citenamefont {Wang},
  \citenamefont {Zhao},\ and\ \citenamefont {Wang}}]{Gui2020}%
  \BibitemOpen
  \bibfield  {author} {\bibinfo {author} {\bibfnamefont {R.}~\bibnamefont
  {Gui}}, \bibinfo {author} {\bibfnamefont {Z.}~\bibnamefont {Xue}}, \bibinfo
  {author} {\bibfnamefont {X.}~\bibnamefont {Zhou}}, \bibinfo {author}
  {\bibfnamefont {C.}~\bibnamefont {Gu}}, \bibinfo {author} {\bibfnamefont
  {X.}~\bibnamefont {Ren}}, \bibinfo {author} {\bibfnamefont {H.}~\bibnamefont
  {Cheng}}, \bibinfo {author} {\bibfnamefont {D.}~\bibnamefont {Ma}}, \bibinfo
  {author} {\bibfnamefont {J.}~\bibnamefont {Qin}}, \bibinfo {author}
  {\bibfnamefont {Y.}~\bibnamefont {Liang}}, \bibinfo {author} {\bibfnamefont
  {X.}~\bibnamefont {Yan}}, \bibinfo {author} {\bibfnamefont {J.}~\bibnamefont
  {Zhang}}, \bibinfo {author} {\bibfnamefont {X.}~\bibnamefont {Zhang}},
  \bibinfo {author} {\bibfnamefont {X.}~\bibnamefont {Yu}}, \bibinfo {author}
  {\bibfnamefont {L.}~\bibnamefont {Wang}}, \bibinfo {author} {\bibfnamefont
  {Y.}~\bibnamefont {Zhao}},\ and\ \bibinfo {author} {\bibfnamefont
  {S.}~\bibnamefont {Wang}},\ }\bibfield  {title} {\bibinfo {title} {Strain
  stiffening, high load-invariant hardness, and electronic anomalies of boron
  phosphide under pressure},\ }\href
  {https://doi.org/10.1103/PhysRevB.101.035302} {\bibfield  {journal} {\bibinfo
   {journal} {Phys. Rev. B}\ }\textbf {\bibinfo {volume} {101}},\ \bibinfo
  {pages} {035302} (\bibinfo {year} {2020})}\BibitemShut {NoStop}%
\bibitem [{\citenamefont {Plimpton}(1995)}]{PLIMPTON1995}%
  \BibitemOpen
  \bibfield  {author} {\bibinfo {author} {\bibfnamefont {S.}~\bibnamefont
  {Plimpton}},\ }\bibfield  {title} {\bibinfo {title} {Fast parallel algorithms
  for short-range molecular dynamics},\ }\href
  {https://doi.org/https://doi.org/10.1006/jcph.1995.1039} {\bibfield
  {journal} {\bibinfo  {journal} {J. Comput. Phys.}\ }\textbf {\bibinfo
  {volume} {117}},\ \bibinfo {pages} {1} (\bibinfo {year} {1995})}\BibitemShut
  {NoStop}%
\bibitem [{\citenamefont {Wang}\ \emph
  {et~al.}(2018{\natexlab{b}})\citenamefont {Wang}, \citenamefont {Zhang},
  \citenamefont {Han},\ and\ \citenamefont {E}}]{Wang2018}%
  \BibitemOpen
  \bibfield  {author} {\bibinfo {author} {\bibfnamefont {H.}~\bibnamefont
  {Wang}}, \bibinfo {author} {\bibfnamefont {L.}~\bibnamefont {Zhang}},
  \bibinfo {author} {\bibfnamefont {J.}~\bibnamefont {Han}},\ and\ \bibinfo
  {author} {\bibfnamefont {W.}~\bibnamefont {E}},\ }\bibfield  {title}
  {\bibinfo {title} {Deepmd-kit: A deep learning package for many-body
  potential energy representation and molecular dynamics},\ }\href
  {https://doi.org/https://doi.org/10.1016/j.cpc.2018.03.016} {\bibfield
  {journal} {\bibinfo  {journal} {Comput. Phys. Commun.}\ }\textbf {\bibinfo
  {volume} {228}},\ \bibinfo {pages} {178} (\bibinfo {year}
  {2018}{\natexlab{b}})}\BibitemShut {NoStop}%
\bibitem [{\citenamefont {Hirao}\ \emph {et~al.}(2020)\citenamefont {Hirao},
  \citenamefont {Kawaguchi}, \citenamefont {Hirose}, \citenamefont {Shimizu},
  \citenamefont {Ohtani},\ and\ \citenamefont {Ohishi}}]{Hirao2020}%
  \BibitemOpen
  \bibfield  {author} {\bibinfo {author} {\bibfnamefont {N.}~\bibnamefont
  {Hirao}}, \bibinfo {author} {\bibfnamefont {S.~I.}\ \bibnamefont
  {Kawaguchi}}, \bibinfo {author} {\bibfnamefont {K.}~\bibnamefont {Hirose}},
  \bibinfo {author} {\bibfnamefont {K.}~\bibnamefont {Shimizu}}, \bibinfo
  {author} {\bibfnamefont {E.}~\bibnamefont {Ohtani}},\ and\ \bibinfo {author}
  {\bibfnamefont {Y.}~\bibnamefont {Ohishi}},\ }\bibfield  {title} {\bibinfo
  {title} {New developments in high-pressure x-ray diffraction beamline for
  diamond anvil cell at spring-8},\ }\href {https://doi.org/10.1063/1.5126038}
  {\bibfield  {journal} {\bibinfo  {journal} {Matter Radiat. at Extremes}\
  }\textbf {\bibinfo {volume} {5}},\ \bibinfo {pages} {018403} (\bibinfo {year}
  {2020})}\BibitemShut {NoStop}%
\bibitem [{\citenamefont {Hammersley}\ \emph {et~al.}(1996)\citenamefont
  {Hammersley}, \citenamefont {Svensson}, \citenamefont {Hanfland},
  \citenamefont {Fitch},\ and\ \citenamefont {Hausermann}}]{Hammersley1996}%
  \BibitemOpen
  \bibfield  {author} {\bibinfo {author} {\bibfnamefont {A.~P.}\ \bibnamefont
  {Hammersley}}, \bibinfo {author} {\bibfnamefont {S.~O.}\ \bibnamefont
  {Svensson}}, \bibinfo {author} {\bibfnamefont {M.}~\bibnamefont {Hanfland}},
  \bibinfo {author} {\bibfnamefont {A.~N.}\ \bibnamefont {Fitch}},\ and\
  \bibinfo {author} {\bibfnamefont {D.}~\bibnamefont {Hausermann}},\ }\bibfield
   {title} {\bibinfo {title} {Two-dimensional detector software: From real
  detector to idealised image or two-theta scan},\ }\href
  {https://doi.org/10.1080/08957959608201408} {\bibfield  {journal} {\bibinfo
  {journal} {High Pressure Res.}\ }\textbf {\bibinfo {volume} {14}},\ \bibinfo
  {pages} {235} (\bibinfo {year} {1996})}\BibitemShut {NoStop}%
\bibitem [{\citenamefont {Walker}\ \emph {et~al.}(2002)\citenamefont {Walker},
  \citenamefont {Cranswick}, \citenamefont {Verma}, \citenamefont {Clark},\
  and\ \citenamefont {Buhre}}]{Walker2002}%
  \BibitemOpen
  \bibfield  {author} {\bibinfo {author} {\bibfnamefont {D.}~\bibnamefont
  {Walker}}, \bibinfo {author} {\bibfnamefont {L.~M.}\ \bibnamefont
  {Cranswick}}, \bibinfo {author} {\bibfnamefont {P.~K.}\ \bibnamefont
  {Verma}}, \bibinfo {author} {\bibfnamefont {S.~M.}\ \bibnamefont {Clark}},\
  and\ \bibinfo {author} {\bibfnamefont {S.}~\bibnamefont {Buhre}},\ }\bibfield
   {title} {\bibinfo {title} {Thermal equations of state for b1 and b2 kcl},\
  }\href {https://doi.org/doi:10.2138/am-2002-0701} {\bibfield  {journal}
  {\bibinfo  {journal} {Am. Mineral.}\ }\textbf {\bibinfo {volume} {87}},\
  \bibinfo {pages} {805} (\bibinfo {year} {2002})}\BibitemShut {NoStop}%
\bibitem [{\citenamefont {Kresse}\ and\ \citenamefont
  {Furthm\"uller}(1996)}]{Kresse1996}%
  \BibitemOpen
  \bibfield  {author} {\bibinfo {author} {\bibfnamefont {G.}~\bibnamefont
  {Kresse}}\ and\ \bibinfo {author} {\bibfnamefont {J.}~\bibnamefont
  {Furthm\"uller}},\ }\bibfield  {title} {\bibinfo {title} {Efficient iterative
  schemes for ab initio total-energy calculations using a plane-wave basis
  set},\ }\href {https://doi.org/10.1103/PhysRevB.54.11169} {\bibfield
  {journal} {\bibinfo  {journal} {Phys. Rev. B}\ }\textbf {\bibinfo {volume}
  {54}},\ \bibinfo {pages} {11169} (\bibinfo {year} {1996})}\BibitemShut
  {NoStop}%
\bibitem [{\citenamefont {Bl\"ochl}(1994)}]{Blochl1994}%
  \BibitemOpen
  \bibfield  {author} {\bibinfo {author} {\bibfnamefont {P.~E.}\ \bibnamefont
  {Bl\"ochl}},\ }\bibfield  {title} {\bibinfo {title} {Projector augmented-wave
  method},\ }\href {https://doi.org/10.1103/PhysRevB.50.17953} {\bibfield
  {journal} {\bibinfo  {journal} {Phys. Rev. B}\ }\textbf {\bibinfo {volume}
  {50}},\ \bibinfo {pages} {17953} (\bibinfo {year} {1994})}\BibitemShut
  {NoStop}%
\bibitem [{\citenamefont {Kresse}\ and\ \citenamefont
  {Joubert}(1999)}]{Kresse1999}%
  \BibitemOpen
  \bibfield  {author} {\bibinfo {author} {\bibfnamefont {G.}~\bibnamefont
  {Kresse}}\ and\ \bibinfo {author} {\bibfnamefont {D.}~\bibnamefont
  {Joubert}},\ }\bibfield  {title} {\bibinfo {title} {From ultrasoft
  pseudopotentials to the projector augmented-wave method},\ }\href
  {https://doi.org/10.1103/PhysRevB.59.1758} {\bibfield  {journal} {\bibinfo
  {journal} {Phys. Rev. B}\ }\textbf {\bibinfo {volume} {59}},\ \bibinfo
  {pages} {1758} (\bibinfo {year} {1999})}\BibitemShut {NoStop}%
\bibitem [{\citenamefont {Perdew}\ \emph {et~al.}(1996)\citenamefont {Perdew},
  \citenamefont {Burke},\ and\ \citenamefont {Ernzerhof}}]{Perdew1996}%
  \BibitemOpen
  \bibfield  {author} {\bibinfo {author} {\bibfnamefont {J.~P.}\ \bibnamefont
  {Perdew}}, \bibinfo {author} {\bibfnamefont {K.}~\bibnamefont {Burke}},\ and\
  \bibinfo {author} {\bibfnamefont {M.}~\bibnamefont {Ernzerhof}},\ }\bibfield
  {title} {\bibinfo {title} {Generalized gradient approximation made simple},\
  }\href {https://doi.org/10.1103/PhysRevLett.77.3865} {\bibfield  {journal}
  {\bibinfo  {journal} {Phys. Rev. Lett.}\ }\textbf {\bibinfo {volume} {77}},\
  \bibinfo {pages} {3865} (\bibinfo {year} {1996})}\BibitemShut {NoStop}%
\bibitem [{\citenamefont {Monkhorst}\ and\ \citenamefont
  {Pack}(1976)}]{Monkhorst1976}%
  \BibitemOpen
  \bibfield  {author} {\bibinfo {author} {\bibfnamefont {H.~J.}\ \bibnamefont
  {Monkhorst}}\ and\ \bibinfo {author} {\bibfnamefont {J.~D.}\ \bibnamefont
  {Pack}},\ }\bibfield  {title} {\bibinfo {title} {Special points for
  brillouin-zone integrations},\ }\href
  {https://doi.org/10.1103/PhysRevB.13.5188} {\bibfield  {journal} {\bibinfo
  {journal} {Phys. Rev. B}\ }\textbf {\bibinfo {volume} {13}},\ \bibinfo
  {pages} {5188} (\bibinfo {year} {1976})}\BibitemShut {NoStop}%
\bibitem [{\citenamefont {Oganov}\ and\ \citenamefont
  {Glass}(2006)}]{Oganov2006}%
  \BibitemOpen
  \bibfield  {author} {\bibinfo {author} {\bibfnamefont {A.~R.}\ \bibnamefont
  {Oganov}}\ and\ \bibinfo {author} {\bibfnamefont {C.~W.}\ \bibnamefont
  {Glass}},\ }\bibfield  {title} {\bibinfo {title} {Crystal structure
  prediction using ab initio evolutionary techniques: Principles and
  applications},\ }\href {https://doi.org/10.1063/1.2210932} {\bibfield
  {journal} {\bibinfo  {journal} {J. Chem. Phys.}\ }\textbf {\bibinfo {volume}
  {124}},\ \bibinfo {pages} {244704} (\bibinfo {year} {2006})}\BibitemShut
  {NoStop}%
\bibitem [{\citenamefont {Lyakhov}\ \emph {et~al.}(2013)\citenamefont
  {Lyakhov}, \citenamefont {Oganov}, \citenamefont {Stokes},\ and\
  \citenamefont {Zhu}}]{Lyakhov2013}%
  \BibitemOpen
  \bibfield  {author} {\bibinfo {author} {\bibfnamefont {A.~O.}\ \bibnamefont
  {Lyakhov}}, \bibinfo {author} {\bibfnamefont {A.~R.}\ \bibnamefont {Oganov}},
  \bibinfo {author} {\bibfnamefont {H.~T.}\ \bibnamefont {Stokes}},\ and\
  \bibinfo {author} {\bibfnamefont {Q.}~\bibnamefont {Zhu}},\ }\bibfield
  {title} {\bibinfo {title} {New developments in evolutionary structure
  prediction algorithm uspex},\ }\href
  {https://doi.org/https://doi.org/10.1016/j.cpc.2012.12.009} {\bibfield
  {journal} {\bibinfo  {journal} {Comput. Phys. Commun.}\ }\textbf {\bibinfo
  {volume} {184}},\ \bibinfo {pages} {1172 } (\bibinfo {year}
  {2013})}\BibitemShut {NoStop}%
\bibitem [{\citenamefont {Kamaeva}\ \emph {et~al.}(2020)\citenamefont
  {Kamaeva}, \citenamefont {Ryltsev}, \citenamefont {Suslov},\ and\
  \citenamefont {Chtchelkatchev}}]{Kamaeva2020}%
  \BibitemOpen
  \bibfield  {author} {\bibinfo {author} {\bibfnamefont {L.~V.}\ \bibnamefont
  {Kamaeva}}, \bibinfo {author} {\bibfnamefont {R.~E.}\ \bibnamefont
  {Ryltsev}}, \bibinfo {author} {\bibfnamefont {A.~A.}\ \bibnamefont
  {Suslov}},\ and\ \bibinfo {author} {\bibfnamefont {N.~M.}\ \bibnamefont
  {Chtchelkatchev}},\ }\bibfield  {title} {\bibinfo {title} {Effect of copper
  concentration on the structure and properties of
  al{\textendash}cu{\textendash}fe and al{\textendash}cu{\textendash}ni
  melts},\ }\href {https://doi.org/10.1088/1361-648x/ab73a6} {\bibfield
  {journal} {\bibinfo  {journal} {J. Phys.: Condens. Matter}\ }\textbf
  {\bibinfo {volume} {32}},\ \bibinfo {pages} {224003} (\bibinfo {year}
  {2020})}\BibitemShut {NoStop}%
\bibitem [{\citenamefont {Ryltsev}\ and\ \citenamefont
  {Chtchelkatchev}(2022)}]{ryltsev2022}%
  \BibitemOpen
  \bibfield  {author} {\bibinfo {author} {\bibfnamefont {R.}~\bibnamefont
  {Ryltsev}}\ and\ \bibinfo {author} {\bibfnamefont {N.}~\bibnamefont
  {Chtchelkatchev}},\ }\bibfield  {title} {\bibinfo {title} {Deep machine
  learning potentials for multicomponent metallic melts: Development,
  predictability and compositional transferability},\ }\href
  {https://doi.org/https://doi.org/10.1016/j.molliq.2021.118181} {\bibfield
  {journal} {\bibinfo  {journal} {J. Mol. Liq.}\ }\textbf {\bibinfo {volume}
  {349}},\ \bibinfo {pages} {118181} (\bibinfo {year} {2022})}\BibitemShut
  {NoStop}%
\bibitem [{\citenamefont {Kondratyuk}\ \emph {et~al.}(2023)\citenamefont
  {Kondratyuk}, \citenamefont {Ryltsev}, \citenamefont {Ankudinov},\ and\
  \citenamefont {Chtchelkatchev}}]{Kondratyuk2023JML}%
  \BibitemOpen
  \bibfield  {author} {\bibinfo {author} {\bibfnamefont {N.}~\bibnamefont
  {Kondratyuk}}, \bibinfo {author} {\bibfnamefont {R.}~\bibnamefont {Ryltsev}},
  \bibinfo {author} {\bibfnamefont {V.}~\bibnamefont {Ankudinov}},\ and\
  \bibinfo {author} {\bibfnamefont {N.}~\bibnamefont {Chtchelkatchev}},\
  }\bibfield  {title} {\bibinfo {title} {First-principles calculations of the
  viscosity in multicomponent metallic melts: Al-cu-ni as a test case},\ }\href
  {https://doi.org/https://doi.org/10.1016/j.molliq.2023.121751} {\bibfield
  {journal} {\bibinfo  {journal} {Journal of Molecular Liquids}\ }\textbf
  {\bibinfo {volume} {380}},\ \bibinfo {pages} {121751} (\bibinfo {year}
  {2023})}\BibitemShut {NoStop}%
\bibitem [{\citenamefont {Zhang}\ \emph {et~al.}(2020)\citenamefont {Zhang},
  \citenamefont {Wang}, \citenamefont {Chen}, \citenamefont {Zeng},
  \citenamefont {Zhang}, \citenamefont {Wang},\ and\ \citenamefont
  {E}}]{Zhang2020}%
  \BibitemOpen
  \bibfield  {author} {\bibinfo {author} {\bibfnamefont {Y.}~\bibnamefont
  {Zhang}}, \bibinfo {author} {\bibfnamefont {H.}~\bibnamefont {Wang}},
  \bibinfo {author} {\bibfnamefont {W.}~\bibnamefont {Chen}}, \bibinfo {author}
  {\bibfnamefont {J.}~\bibnamefont {Zeng}}, \bibinfo {author} {\bibfnamefont
  {L.}~\bibnamefont {Zhang}}, \bibinfo {author} {\bibfnamefont
  {H.}~\bibnamefont {Wang}},\ and\ \bibinfo {author} {\bibfnamefont
  {W.}~\bibnamefont {E}},\ }\bibfield  {title} {\bibinfo {title} {Dp-gen: A
  concurrent learning platform for the generation of reliable deep learning
  based potential energy models},\ }\href
  {https://doi.org/https://doi.org/10.1016/j.cpc.2020.107206} {\bibfield
  {journal} {\bibinfo  {journal} {Comput. Phys. Commun.}\ }\textbf {\bibinfo
  {volume} {253}},\ \bibinfo {pages} {107206} (\bibinfo {year}
  {2020})}\BibitemShut {NoStop}%
\bibitem [{\citenamefont {Kurakevych}\ \emph {et~al.}(2015)\citenamefont
  {Kurakevych}, \citenamefont {Godec},\ and\ \citenamefont
  {Solozhenko}}]{Kurakevych2015}%
  \BibitemOpen
  \bibfield  {author} {\bibinfo {author} {\bibfnamefont {O.~O.}\ \bibnamefont
  {Kurakevych}}, \bibinfo {author} {\bibfnamefont {Y.~L.}\ \bibnamefont
  {Godec}},\ and\ \bibinfo {author} {\bibfnamefont {V.~L.}\ \bibnamefont
  {Solozhenko}},\ }\bibfield  {title} {\bibinfo {title} {Equations of state of
  novel solids synthesized under extreme pressure{\textendash}temperature
  conditions},\ }\href {https://doi.org/10.1088/1742-6596/653/1/012080}
  {\bibfield  {journal} {\bibinfo  {journal} {J. Phys: Conf. Ser.}\ }\textbf
  {\bibinfo {volume} {653}},\ \bibinfo {pages} {012080} (\bibinfo {year}
  {2015})}\BibitemShut {NoStop}%
\bibitem [{\citenamefont {Vega}\ \emph {et~al.}(2008)\citenamefont {Vega},
  \citenamefont {Sanz}, \citenamefont {Abascal},\ and\ \citenamefont
  {Noya}}]{Vega2008}%
  \BibitemOpen
  \bibfield  {author} {\bibinfo {author} {\bibfnamefont {C.}~\bibnamefont
  {Vega}}, \bibinfo {author} {\bibfnamefont {E.}~\bibnamefont {Sanz}}, \bibinfo
  {author} {\bibfnamefont {J.~L.~F.}\ \bibnamefont {Abascal}},\ and\ \bibinfo
  {author} {\bibfnamefont {E.~G.}\ \bibnamefont {Noya}},\ }\bibfield  {title}
  {\bibinfo {title} {Determination of phase diagrams via computer simulation:
  methodology and applications to water, electrolytes and proteins},\ }\href
  {https://doi.org/10.1088/0953-8984/20/15/153101} {\bibfield  {journal}
  {\bibinfo  {journal} {J. Phys.: Condens. Matter}\ }\textbf {\bibinfo {volume}
  {20}},\ \bibinfo {pages} {153101} (\bibinfo {year} {2008})}\BibitemShut
  {NoStop}%
\bibitem [{\citenamefont {Zeng}(2020)}]{dpti2020}%
  \BibitemOpen
  \bibfield  {author} {\bibinfo {author} {\bibfnamefont {J.}~\bibnamefont
  {Zeng}},\ }\href {https://github.com/deepmodeling/dpti} {}\bibinfo
  {howpublished} {\url{https://github.com/deepmodeling/dpti}} (\bibinfo {year}
  {2020})\BibitemShut {NoStop}%
\bibitem [{\citenamefont {Zou}\ \emph {et~al.}(2020)\citenamefont {Zou},
  \citenamefont {Xiang},\ and\ \citenamefont {Dai}}]{Zou2020}%
  \BibitemOpen
  \bibfield  {author} {\bibinfo {author} {\bibfnamefont {Y.}~\bibnamefont
  {Zou}}, \bibinfo {author} {\bibfnamefont {S.}~\bibnamefont {Xiang}},\ and\
  \bibinfo {author} {\bibfnamefont {C.}~\bibnamefont {Dai}},\ }\bibfield
  {title} {\bibinfo {title} {Investigation on the efficiency and accuracy of
  methods for calculating melting temperature by molecular dynamics
  simulation},\ }\href {https://doi.org/10.1016/j.commatsci.2019.109156}
  {\bibfield  {journal} {\bibinfo  {journal} {Comput. Mater. Sci.}\ }\textbf
  {\bibinfo {volume} {171}},\ \bibinfo {pages} {109156} (\bibinfo {year}
  {2020})}\BibitemShut {NoStop}%
\bibitem [{\citenamefont {Rozas}\ \emph {et~al.}(2022)\citenamefont {Rozas},
  \citenamefont {Ankudinov},\ and\ \citenamefont {Galenko}}]{Rozas2022}%
  \BibitemOpen
  \bibfield  {author} {\bibinfo {author} {\bibfnamefont {R.}~\bibnamefont
  {Rozas}}, \bibinfo {author} {\bibfnamefont {V.}~\bibnamefont {Ankudinov}},\
  and\ \bibinfo {author} {\bibfnamefont {P.}~\bibnamefont {Galenko}},\
  }\bibfield  {title} {\bibinfo {title} {Kinetics of rapid growth and melting
  of alni alloying crystals: phase field theory versus atomistic simulations
  revisited},\ }\href {https://doi.org/10.1088/1361-648x/ac9a1c} {\bibfield
  {journal} {\bibinfo  {journal} {J. Phys.: Condens. Matter}\ }\textbf
  {\bibinfo {volume} {34}},\ \bibinfo {pages} {494002} (\bibinfo {year}
  {2022})}\BibitemShut {NoStop}%
\bibitem [{\citenamefont {Sun}\ \emph {et~al.}(2022)\citenamefont {Sun},
  \citenamefont {Qian}, \citenamefont {Pang}, \citenamefont {Lu}, \citenamefont
  {Guo},\ and\ \citenamefont {Wang}}]{Sun2022}%
  \BibitemOpen
  \bibfield  {author} {\bibinfo {author} {\bibfnamefont {Y.}~\bibnamefont
  {Sun}}, \bibinfo {author} {\bibfnamefont {G.}~\bibnamefont {Qian}}, \bibinfo
  {author} {\bibfnamefont {S.}~\bibnamefont {Pang}}, \bibinfo {author}
  {\bibfnamefont {J.}~\bibnamefont {Lu}}, \bibinfo {author} {\bibfnamefont
  {J.}~\bibnamefont {Guo}},\ and\ \bibinfo {author} {\bibfnamefont
  {Z.}~\bibnamefont {Wang}},\ }\bibfield  {title} {\bibinfo {title} {Partition
  model for trace elements between liquid metal and silicate melts involving
  the interfacial transition structure: An exploratory two-phase
  first-principles molecular dynamics study},\ }\href
  {https://doi.org/10.1016/j.molliq.2022.120048} {\bibfield  {journal}
  {\bibinfo  {journal} {J. Mol. Liq.}\ }\textbf {\bibinfo {volume} {364}},\
  \bibinfo {pages} {120048} (\bibinfo {year} {2022})}\BibitemShut {NoStop}%
\bibitem [{\citenamefont {Price}\ \emph {et~al.}(2009)\citenamefont {Price},
  \citenamefont {Alatas}, \citenamefont {Hennet}, \citenamefont {Jakse},
  \citenamefont {Krishnan}, \citenamefont {Pasturel}, \citenamefont
  {Pozdnyakova}, \citenamefont {Saboungi}, \citenamefont {Said}, \citenamefont
  {Scheunemann}, \citenamefont {Schirmacher},\ and\ \citenamefont
  {Sinn}}]{Price2009PRL}%
  \BibitemOpen
  \bibfield  {author} {\bibinfo {author} {\bibfnamefont {D.~L.}\ \bibnamefont
  {Price}}, \bibinfo {author} {\bibfnamefont {A.}~\bibnamefont {Alatas}},
  \bibinfo {author} {\bibfnamefont {L.}~\bibnamefont {Hennet}}, \bibinfo
  {author} {\bibfnamefont {N.}~\bibnamefont {Jakse}}, \bibinfo {author}
  {\bibfnamefont {S.}~\bibnamefont {Krishnan}}, \bibinfo {author}
  {\bibfnamefont {A.}~\bibnamefont {Pasturel}}, \bibinfo {author}
  {\bibfnamefont {I.}~\bibnamefont {Pozdnyakova}}, \bibinfo {author}
  {\bibfnamefont {M.-L.}\ \bibnamefont {Saboungi}}, \bibinfo {author}
  {\bibfnamefont {A.}~\bibnamefont {Said}}, \bibinfo {author} {\bibfnamefont
  {R.}~\bibnamefont {Scheunemann}}, \bibinfo {author} {\bibfnamefont
  {W.}~\bibnamefont {Schirmacher}},\ and\ \bibinfo {author} {\bibfnamefont
  {H.}~\bibnamefont {Sinn}},\ }\bibfield  {title} {\bibinfo {title} {Liquid
  boron: X-ray measurements and ab initio molecular dynamics simulations},\
  }\href {https://doi.org/10.1103/PhysRevB.79.134201} {\bibfield  {journal}
  {\bibinfo  {journal} {Phys. Rev. B}\ }\textbf {\bibinfo {volume} {79}},\
  \bibinfo {pages} {134201} (\bibinfo {year} {2009})}\BibitemShut {NoStop}%
\bibitem [{\citenamefont {Durandurdu}(2015)}]{Durandurdu2015JNCS}%
  \BibitemOpen
  \bibfield  {author} {\bibinfo {author} {\bibfnamefont {M.}~\bibnamefont
  {Durandurdu}},\ }\bibfield  {title} {\bibinfo {title} {Liquid boron and
  amorphous boron: An ab initio molecular dynamics study},\ }\href
  {https://doi.org/10.1016/j.jnoncrysol.2015.03.004} {\bibfield  {journal}
  {\bibinfo  {journal} {Journal of Non-Crystalline Solids}\ }\textbf {\bibinfo
  {volume} {417-418}},\ \bibinfo {pages} {10} (\bibinfo {year}
  {2015})}\BibitemShut {NoStop}%
\bibitem [{\citenamefont {Senda}\ \emph {et~al.}(2002)\citenamefont {Senda},
  \citenamefont {Shimojo},\ and\ \citenamefont {Hoshino}}]{Senda2002JPCM}%
  \BibitemOpen
  \bibfield  {author} {\bibinfo {author} {\bibfnamefont {Y.}~\bibnamefont
  {Senda}}, \bibinfo {author} {\bibfnamefont {F.}~\bibnamefont {Shimojo}},\
  and\ \bibinfo {author} {\bibfnamefont {K.}~\bibnamefont {Hoshino}},\
  }\bibfield  {title} {\bibinfo {title} {The metal-nonmetal transition of
  liquid phosphorus by ab initio molecular-dynamics simulations},\ }\href
  {https://doi.org/10.1088/0953-8984/14/14/304} {\bibfield  {journal} {\bibinfo
   {journal} {Journal of Physics: Condensed Matter}\ }\textbf {\bibinfo
  {volume} {14}},\ \bibinfo {pages} {3715} (\bibinfo {year}
  {2002})}\BibitemShut {NoStop}%
\bibitem [{\citenamefont {Ghiringhelli}\ and\ \citenamefont
  {Meijer}(2005)}]{Ghiringhelli2005JCP}%
  \BibitemOpen
  \bibfield  {author} {\bibinfo {author} {\bibfnamefont {L.~M.}\ \bibnamefont
  {Ghiringhelli}}\ and\ \bibinfo {author} {\bibfnamefont {E.~J.}\ \bibnamefont
  {Meijer}},\ }\bibfield  {title} {\bibinfo {title} {{Phosphorus: First
  principle simulation of a liquid–liquid phase transition}},\ }\bibfield
  {journal} {\bibinfo  {journal} {The Journal of Chemical Physics}\ }\textbf
  {\bibinfo {volume} {122}},\ \href {https://doi.org/10.1063/1.1895717}
  {10.1063/1.1895717} (\bibinfo {year} {2005})\BibitemShut {NoStop}%
\bibitem [{\citenamefont {Porowski}\ \emph {et~al.}(2019)\citenamefont
  {Porowski}, \citenamefont {Sadovyi}, \citenamefont {Karbovnyk}, \citenamefont
  {Gierlotka}, \citenamefont {Rzoska}, \citenamefont {Petrusha}, \citenamefont
  {Stratiichuk}, \citenamefont {Turkevich},\ and\ \citenamefont
  {Grzegory}}]{Porowski2019JCrystGrowth}%
  \BibitemOpen
  \bibfield  {author} {\bibinfo {author} {\bibfnamefont {S.}~\bibnamefont
  {Porowski}}, \bibinfo {author} {\bibfnamefont {B.}~\bibnamefont {Sadovyi}},
  \bibinfo {author} {\bibfnamefont {I.}~\bibnamefont {Karbovnyk}}, \bibinfo
  {author} {\bibfnamefont {S.}~\bibnamefont {Gierlotka}}, \bibinfo {author}
  {\bibfnamefont {S.}~\bibnamefont {Rzoska}}, \bibinfo {author} {\bibfnamefont
  {I.}~\bibnamefont {Petrusha}}, \bibinfo {author} {\bibfnamefont
  {D.}~\bibnamefont {Stratiichuk}}, \bibinfo {author} {\bibfnamefont
  {V.}~\bibnamefont {Turkevich}},\ and\ \bibinfo {author} {\bibfnamefont
  {I.}~\bibnamefont {Grzegory}},\ }\bibfield  {title} {\bibinfo {title}
  {Melting of tetrahedrally bonded semiconductors: “anomaly” of the phase
  diagram of gan?},\ }\href {https://doi.org/10.1016/j.jcrysgro.2018.09.007}
  {\bibfield  {journal} {\bibinfo  {journal} {Journal of Crystal Growth}\
  }\textbf {\bibinfo {volume} {505}},\ \bibinfo {pages} {5} (\bibinfo {year}
  {2019})}\BibitemShut {NoStop}%
\bibitem [{\citenamefont {Xu}\ \emph {et~al.}(2014)\citenamefont {Xu},
  \citenamefont {Bi}, \citenamefont {Li}, \citenamefont {Wang}, \citenamefont
  {Cao}, \citenamefont {Cai}, \citenamefont {Wang},\ and\ \citenamefont
  {Meng}}]{Xu_2014}%
  \BibitemOpen
  \bibfield  {author} {\bibinfo {author} {\bibfnamefont {L.}~\bibnamefont
  {Xu}}, \bibinfo {author} {\bibfnamefont {Y.}~\bibnamefont {Bi}}, \bibinfo
  {author} {\bibfnamefont {X.}~\bibnamefont {Li}}, \bibinfo {author}
  {\bibfnamefont {Y.}~\bibnamefont {Wang}}, \bibinfo {author} {\bibfnamefont
  {X.}~\bibnamefont {Cao}}, \bibinfo {author} {\bibfnamefont {L.}~\bibnamefont
  {Cai}}, \bibinfo {author} {\bibfnamefont {Z.}~\bibnamefont {Wang}},\ and\
  \bibinfo {author} {\bibfnamefont {C.}~\bibnamefont {Meng}},\ }\bibfield
  {title} {\bibinfo {title} {Phase diagram of tin determined by sound velocity
  measurements on multi-anvil apparatus up to 5{\hspace{0.167em}}{GPa} and
  800{\hspace{0.167em}}k},\ }\href
  {https://doi.org/https://doi.org/10.1063/1.4872458} {\bibfield  {journal}
  {\bibinfo  {journal} {Journal of Applied Physics}\ }\textbf {\bibinfo
  {volume} {115}},\ \bibinfo {pages} {164903} (\bibinfo {year}
  {2014})}\BibitemShut {NoStop}%
\bibitem [{\citenamefont {Dorner}\ \emph {et~al.}(2018)\citenamefont {Dorner},
  \citenamefont {Sukurma}, \citenamefont {Dellago},\ and\ \citenamefont
  {Kresse}}]{Dorner2018}%
  \BibitemOpen
  \bibfield  {author} {\bibinfo {author} {\bibfnamefont {F.}~\bibnamefont
  {Dorner}}, \bibinfo {author} {\bibfnamefont {Z.}~\bibnamefont {Sukurma}},
  \bibinfo {author} {\bibfnamefont {C.}~\bibnamefont {Dellago}},\ and\ \bibinfo
  {author} {\bibfnamefont {G.}~\bibnamefont {Kresse}},\ }\bibfield  {title}
  {\bibinfo {title} {Melting si: Beyond density functional theory},\ }\bibfield
   {journal} {\bibinfo  {journal} {Physical Review Letters}\ }\textbf {\bibinfo
  {volume} {121}},\ \href {https://doi.org/10.1103/physrevlett.121.195701}
  {10.1103/physrevlett.121.195701} (\bibinfo {year} {2018})\BibitemShut
  {NoStop}%
\end{thebibliography}%
\end{document}